\documentclass[graybox]{svmult}

\usepackage{type1cm}        
\usepackage{makeidx}         
\usepackage{graphicx}        
\usepackage{multicol}        
\usepackage[bottom]{footmisc}

\usepackage{newtxtext}       %
\usepackage[varvw]{newtxmath}       
\usepackage{tensor}
\usepackage{braket}
\usepackage{cleveref}
\usepackage{mdframed}
\newenvironment{B0x}[2][Box]
    { \begin{mdframed}[backgroundcolor=gray!20] \textbf{#1 #2} }
    {  \end{mdframed}}
\makeindex             
\def\be{\begin{equation}}
    \def\ee{\end{equation}}
    \def\bea{\begin{eqnarray}}
    \def\eea{\end{eqnarray}}
    \def\pd{\partial}
    \def\a{\alpha}
    \def\b{\beta}
    \def\g{\gamma}
    
    \def\d{\delta}
    \def\m{\mu}
    \def\n{\nu}
    \def\t{\tau}

    \def\l{\lambda}

    \def\r{\rho}

    \def\bR{\bar{R}}

    \def\bn{\bar{\nabla}}
    \def\bR{\bar{R}}

    \def\s{\sigma}

    \def\e{\epsilon}
    \def\bi{\begin{itemize}}
    \def\ei{\end{itemize}}
    \def\bg{\bar{g}}

    \usepackage[compat=1.1.0]{tikz-feynman} 
    \usetikzlibrary{patterns,snakes}
    \tikzstyle{spring}=[line width=0.8,blue!7!black!80,snake=coil,segment amplitude=4.25,segment length=4.75,line cap=round]

\begin{document}	
	
\title*{A Primer on Unimodular Gravity\thanks{Invited chapter for {\it Handbook of Quantum Gravity} (Eds. C.~Bambi, L.~Modesto, and I.~L.~Shapiro, Springer 2023).}}
\author{Enrique Álvarez and Eduardo Velasco-Aja}
\institute{Enrique Álvarez \at IFT-UAM/CSIC, Cantoblanco, Madrid, \email{enrique.alvarez@uam.es}
\and Eduardo Velasco-Aja \at IFT-UAM/CSIC, Cantoblanco, Madrid \email{eduardo.velasco@uam.es}}
%
%
\maketitle

\abstract*{This chapter provides an introduction to Unimodular Gravity both at the classical and quantum level, discussing the r\^ole it might play in the partial solution of the Cosmological Constant problem. 
The main objective of this work is to serve as a conceptual introduction to Unimodular gravity without disregarding computational detail. In that sense, techniques used at the research level are presented and applied in detail. 
}
\abstract{This chapter provides an introduction to Unimodular Gravity both at the classical and quantum level, discussing the r\^ole it might play in the partial solution of the Cosmological Constant problem. 
The main objective of this work is to serve as a conceptual introduction to Unimodular gravity without disregarding computational detail. In that sense, techniques used at the research level are presented and applied in detail. 
}
 \keywords{Unimodular Gravity; Cosmological Constant Problem; Modified Gravity.}
 \section{Introduction.}\label{sec:intro}
 Unimodular gravity (UG) was first considered by Einstein in 1919 \cite{Einstein1919} as an attempt to make a connection between gravity and Mie's theory of electromagnetic wave scattering, published a few years prior.
 The original proposal was a far cry from the current status of UG. 
 However, seven years later, in his book \cite{Pauli1981}, Pauli discussed the topic in a much more modern flavor.
The theory took rise when the attractive properties of UG in the context of a field theory were noticed in \cite{VanderBij}, which was soon followed by the works \cite{Henneaux1989, Unruh1989, Bunchmuller1988, Bunchmuller1989}.
 \\
 UG was originally conceived as a metric theory of gravity in which the determinant of the metric satisfies the unimodular condition; 
 \begin{equation}
 g\equiv|\det(\tensor{g}{_\mu _\nu})|=1.
 \end{equation}
 The general condition for a diffeomorphism (\textit{Diff} in the sequel) 
 \be
 x\rightarrow x^\prime,
 \ee
 to preserve the unimodular condition is that the jacobian satisfies
 \begin{equation}
 J\equiv \det\,\frac{\pd \tensor{x}{^\prime^\lambda} }{\pd x^\a}=\pm 1.
 \end{equation}
 In a more modern language, UG is often introduced by considering an action functional for the metric and the matter fields, represented here by $\phi$, 
 \begin{equation}
 S[g,\phi]\equiv\int_{}^{} d(vol)\,\mathcal{L}[g,\phi],
 \end{equation}
 for which the volume form is fixed, i.e.,
 \begin{equation}
 d(vol)=\omega \tensor{\epsilon}{_{\mu_1}_\dots_{\mu_d}}dx^{\mu_1}\wedge\cdots\wedge dx^{\mu_d} ,
 \end{equation}
one then sets, $\sqrt{g}=\omega$.
 This theory then will not be invariant under general coordinate transformations \textit{Diff} but under those coordinate transformations that preserve the determinant of the metric (and hence the volume form). 
 These are the volume preserving or transverse \textit{Diffs}; \textit{VPD}. 
 
 It is well known \cite{Ismagilov} that not all \textit{Diffs} that are close to the identity can be obtained by the exponential map; nevertheless, the ones for which this is possible can be represented by a transverse vector
 \be
 \nabla_\m \xi^\m=0.
 \ee
 It is quite easy to check that VPD generate a subgroup of all \textit{Diffs} connected to the identity, $\mbox{\textit{Diff}}$, because 
 if $\xi_1^\m$ and $\xi_2^\m$ are transverse, then so is their commutator, 
 \be
 \left[\xi_1^\m,\xi_2^\m\right] \in VPD\, .
 \ee
 \\
 In UG, the equations of motion (EM)
 \be
 E_{\m\n}\equiv {\d S\over \d g^{\m\n}},
 \ee
 obey, see \cite{Alvarez2005}, the following condition;
 \be
 \nabla_{[\lambda|} \nabla^\m E_{\m|\nu]}=0, 
 \ee
 which is equivalent to the existence of a function $S$ such that,
 \begin{equation}
   \nabla_\m E^\m\,_\n=\nabla_\n S,
 \end{equation}
 as long as the de Rham cohomology of the manifold is trivial $H_{dR}^3(M)=0$; otherwise, harmonic contributions have to be included. In fact
 \begin{align}
 \d S&= \int d(vol) E_{\m\n} \left(\nabla^\m \xi^\n+\nabla^\n \xi^\m\right)=\nonumber\\
 &=\int d(vol) E_{\m\n}\left(\nabla^\m \eta^{\n\a\b\g} \nabla_\a \Omega_{\m\g}+\nabla^\n \eta^{\m\a\b\g}\nabla_\a \Omega_{\b\g}\right),
 \end{align}
 where 
 \bea
 &\Omega_{\m\n}=-\Omega_{\n\m},\qquad\eta_{\a\b\g\d}\equiv \sqrt{g}\e_{\a\b\g\d},
 \eea
 and the statement follows. Of course, in the \textit{Diff} invariant case, $S=\text{constant}$.
 \\
Alternatively to this quick presentation, \cref{sec: Linear Field Theory} closely follows the spirit of \cite{VanderBij}. Thus, it reviews the general yet erroneous idea that General Relativity (GR) is the only ghost-free theory of a massless spin-two particle.
\\
 
 Let us precede this construction with a motivation for UG in the context of the Cosmological Constant Problem (CCP); see \cite{Weinber1989} for the original review, or \cite{Carroll2001} for an updated and extensive bibliographic guide of the topic. 
 The aspect of the CCP considered here corresponds to the fact that contrary to Minkowski spacetime, where normal ordering can set vacuum energy to zero, on curved spacetimes, vacuum energy is naively expected to contribute to the stress-energy tensor as
 \begin{equation}
 \braket{\tensor{T}{_\mu^\mu }}_0\tensor{g}{_\mu_\nu},
 \end{equation}
 which has the same form as the CC term on the (EM) of GR,
 \begin{equation}
 \tensor{R}{_\mu_\nu}-\frac{1}{2}R\tensor{g}{_\mu_\nu}-\lambda\tensor{g}{_\mu_\nu}=2 \kappa^2\left(\tensor{\tilde{T}}{_\mu_\nu}+\braket{\tensor{T}{_\mu^\mu }}_0\tensor{g}{_\mu_\nu}\right). 
 \end{equation}
 Here, $\tensor{\tilde{T}}{_\mu_\nu}$ represents all contributions to the stress-energy tensors except the ones coming from the vacuum expectation value of quantum fields.
 
 Therefore in GR, the CC term and vacuum energy contributions to the stress-energy tensor have the same form.
 
 However, any gross estimate for the vacuum energy of quantum fields yields results over 30 orders of magnitude from the observed value for the CC \cite{Weinber1989,Carroll2001}. 
 Thus there is a fine-tuning problem in which the different contributions to the vacuum energy, including those coming from phase transitions, have to cancel to an accuracy way beyond thirty decimal places.
 It is in this sense that the CC is non-natural. 
 
 In \cref{sec:Non-Linear Field Theory}, it is shown that, in the context of UG, the CC appears as an integration constant with no naturalness problem associated. 
 This is followed by a discussion on the all-important problem of coupling sources to UG \cref{sec:Sources}. 
 The status of Birkhoff's theorem in UG is discussed in \cref{sec:Birkhoff}. 
 Section \ref{sec:Cosmology} is an invitation to Unimodular Cosmology.
 In \cref{sec:trees}, tree-level calculations for UG are presented. 
 This is completed by the one-loop effect discussion of \cref{sec:quantization}, where necessary tools to apply the path integral formalism to UG are introduced.

 \section{Linear Field Theory.}\label{sec: Linear Field Theory}
 This section reviews the theory of a linear field that {only propagates a massless spin-two field}, with two polarisation degrees of freedom (DOF).
 As originally discussed in \cite{VanderBij} and later in \cite{Alvarez2006}, here it is shown that the linear \textit{Diff}s invariance, \textit{LDiff}, is too restrictive in the sense that the subgroup of volume-preserving linear diffeomorphisms \textit{LTDiff}, can successfully do the job.
 \\
 The notion of a massless free particle in flat spacetimes is tied to the invariance under the Poincaré group, 
 (in particular to the \textit{proper orthochronous Lorentz group} $ \mathcal{L}^+_\uparrow$), of the unitary representation of the covering group of the \textit{little group } of the four-vector,
 \begin{equation}
 k=\begin{pmatrix}
 E,&0,&0,&E
 \end{pmatrix}.
 \end{equation}
 To make the group finite-dimensional, gauge invariance introduces the equivalence class, 
 \begin{equation}
 \tensor{h}{_\mu_\nu}(k)\equiv \tensor{h}{_\mu_\nu}(k)+2k_{ (\mu} \xi_{ \nu)}(k),
 \end{equation}
 with the two constraints, 
 \begin{equation}
 k^2 \xi_\mu(k)=0\qquad \mbox{and}\qquad k_\mu \xi^\mu(k)=0.
 \end{equation}
 As mentioned in \cite{VanderBij}, the first condition is too restrictive for interaction, so it must be discarded. 
 At this point, if one also drops the second condition, 
 \begin{equation}
 \tensor{h}{_\mu_\nu}(k)\equiv \tensor{h}{_\mu_\nu}(k)+2k_{ (\mu} \xi_{ \nu)}(k),
 \end{equation}
 which is known to correspond to linearized diffeomorphisms \textit{LDiff}. 
 Nevertheless, from the above discussion, one can see that keeping the transverse condition does the job. 
 This would correspond in position space to, 
 \begin{align}
 \tensor{h}{_\mu_\nu}(x)&\equiv \tensor{h}{_\mu_\nu}(x)+2\partial_{ (\mu} \xi_{ \nu)}(x),\label{7} \\
 \partial_\mu \xi^\mu(x)&=0.\label{8} 
 \end{align}
 This characterizes linearized transverse diffeomorphisms, \textit{LTDiff}. 
 
 A neat physical interpretation of \cref{7,8} can be given by realizing that, from a physical perspective, \textit{LTDiff} transformations have three DOF. 
 These three DOF are enough to pass from the five DOF of massive gravity to the two DOF of a theory that only propagates a massless spin-two particle. 
 
 Considering these observations, a linearized lagrangian that is second order in derivatives and such that it is ghost free while only propagating a spin-two massless particle can be built. 
 Writing the more general set of operators that satisfy the above conditions,
 \begin{equation}
 \mathcal{L}\equiv \sum_{i=0}^{4} C_i \mathcal{O}^{(i)},\label{c0} 
 \end{equation}
 where, defining $\mathit{h}\equiv\tensor{\eta}{^\alpha^\beta}\tensor{h}{_\alpha_\beta}=\tensor{h}{_\alpha^\alpha} $ \footnote{When the background metric $\tensor{\bg}{_\alpha_\beta}$ is arbitary, this will generalize to $\mathit{h}\equiv\tensor{\bg}{^\alpha^\beta}\tensor{h}{_\alpha_\beta}. $ } .
 \begin{align}
 &\mathcal{O}^{(1)}\equiv \frac{1}{4}\partial_\mu \tensor{h}{_\alpha_\beta} \partial^\mu \tensor{h}{^\alpha^\beta} && \mathcal{O}^{(2)}\equiv-\frac{1}{2}\partial_\lambda \tensor{h}{^\mu^\lambda}\partial_\rho \tensor{h}{_\mu^\rho}\\
 &\mathcal{O}^{(3)}\equiv \frac{1}{2}\partial_\mu h \partial_\lambda \tensor{h}{^\mu^\lambda} && \mathcal{O}^{(4)}\equiv-\frac{1}{4}\partial^\mu h \partial_\mu h
 \end{align}
 Setting $C_1=1$ for normalization and requiring invariance under \textit{LTDiff} yields\footnote{For Fierz-Pauli $C_i=1$. Thus, \textit{LTDiff} gives a less stringent condition.} 
 \begin{equation}
 C_1=C_2=1.\label{c1} 
 \end{equation}
 Additionally, imposing invariance under Weyl transformations results in, 
 \begin{equation}
 C_3=\frac{2}{n}\qquad \mbox{and }\qquad C_4=\frac{n+2}{n^2}.\label{c2} 
 \end{equation}
 \textit{LWTDiff}\footnote{The reason to consider \textit{LWTDiff} will be discussed in \cref{sec:Non-Linear Field Theory} .} can be obtained combining \cref{c0,c1,c2}.
 
 It is interesting to note that the same constraints for $C_i$ could have been obtained by starting from the Fierz-Pauli Lagrangian (with $C_i=1$) rewriting, 
 \begin{equation}
 \tensor{h}{_\mu_\nu} \to\tensor{h}{_\mu_\nu}-\frac{1}{n}h\, \tensor{\eta}{_\mu_\nu}. \label{c3} 
 \end{equation}
 Note that \cref{c3} {does not represent a field redefinition}, as it is not invertible.
 
 In the next section, the fully non-linear theory is developed, where \textit{LTDiff} is substituted by \textit{TDiff}. 
 Recall that \textit{TDiff} correspond to VPD connected to the identity. 
 
 \section{Non-Linear Field Theory.}\label{sec:Non-Linear Field Theory}
 
 In the previous section, the theory at the linear level was considered. 
The complete, non-linear theory can be obtained by considering the Einstein-Hilbert action for a unimodular metric $\tensor{\gamma}{_\mu_\nu}$, 
 \begin{equation}
 S[\tensor{\hat{g}}{_\mu_\nu}]=-\frac{1}{2\kappa^2}\int d^n\,x\sqrt{\gamma}\,R\left[\tensor{\gamma}{_\mu_\nu}\right].\label{3.1}
 \end{equation}
 The EM can be obtained considering only transverse variations of the metric \cite{Alvarez2005}. 
 Consider for the moment a practical approach that will prove useful when quantizing.
 For this, define an auxiliary field from the unimodular metric,
 \begin{equation}
 \tensor{\gamma}{_\mu_\nu}\rightarrow\tensor{g}{_\mu_\nu}\equiv g^{\frac{1}{n}}\gamma_{\mu\nu}.\label{3.2}
 \end{equation}
 Since \cref{3.1} is invariant only under volume-preserving diffeomorphism, for the allowed changes of coordinates, the determinant of the metric will behave as a scalar and not as a density. 
 Then, \cref{3.2} can be regarded as a conformal (or Weyl) transformation,
 \begin{equation}
 \tensor{\gamma}{_\mu_\nu}\rightarrow\,e^{2 \phi(x)}\tensor{\gamma}{_\mu_\nu}.\label{conformal} 
 \end{equation}
 As previously noted, \cref{3.2} is {\em not} a field redefinition because it is not invertible. 
 Actually, the variation can be written in terms of $ \d \hat{g}_{\m\n}$ 
 \be
 \d \gamma_{\m\n}\equiv M_{\m\n}^{\a\b} \d g_{\a\b}=g^{-{1\over n}}\left({1\over 2}\left(\d^\a_\m\d^\b_\n+\d^\a_\n\d^\b_\m\right)-{1\over n}g^{\a\b} g_{\m\n}\right)\d g_{\a\b}.
 \ee
 It is impossible to recover the generic metric out of Einstein's unimodular metric (the converse is, of course, trivial). 
 This action in an arbitrary frame is invariant under local Weyl transformations of the form of \cref{conformal}.

 The action \eqref{3.1}, after integration by parts\footnote{Usually the integral of a covariant derivative vanishes because it can be written as
 \be
 \bn_\m V^\m={1\over \sqrt{\bg}}\pd_\m\left(\sqrt{\bg}V^\m\right),\nonumber
 \ee
 so that
 \be
 \int d^n x \sqrt{\bg}\bn_\m V^\m=\int d^n x \pd_\m\left(\sqrt{\bg}V^\m\right)=0,\nonumber
 \ee
 assuming vanishing physical effects at the boundary. This is not true anymore with the unimodular measure. What can be written instead is
 
 \be
 \int d^n x \bg^{1\over n}\bn_\m V^\m={n-2\over n}\int d^n x V^\m \bg^{2-n\over n}~\pd_\m \bg.\nonumber
 \ee
 }, reads in terms of $\tensor{g}{_\mu_\nu} $ ,
 \begin{equation}
 S[\tensor{g}{_\mu_\nu}]=-\frac{1}{2\kappa^2}\int d^n\,x\,g^{\frac{1}{n}}\left(R\left[\tensor{g}{_\mu_\nu}\right]+\frac{(n-1)(n-2)}{4n^2}\frac{\nabla_\mu g\nabla^\mu g}{g^2}\right). \label{UGact1} 
 \end{equation}
The gauge group for the  action \eqref{UGact1}, is given, see \cite{Alvarez2016-2}, by the semi-direct product 
 \begin{equation}
 \mbox{WTDiff}= \mbox{Weyl}\ltimes \mbox{TDiff}.
 \end{equation}
 The equations of motion for this action read \footnote{Here, a stress-energy tensor has been introduced for completeness. Note also that $T\equiv\tensor{T}{_\mu_\nu}\tensor{g}{^\mu^\nu} $. Sources are properly dealt with in \cref{sec:Sources}. } \cite{Alvarez2010},
 \begin{align}
 &\tensor{R}{_\mu_\nu}-\frac{1}{n}R\tensor{g}{_\mu_\nu}+\frac{(2-n)(2n-1)}{4n^2}\left(\frac{\nabla_\mu g\nabla_\nu g}{g^2}-\frac{1}{n}\frac{(\nabla g)^2}{g^2}\tensor{g}{_\mu_\nu}\right)+\nonumber\\
 &+\frac{n-2}{2n}\left(\frac{\nabla_\mu\nabla_\nu g}{g}-\frac{1}{n}\frac{\nabla^2g}{g}\tensor{g}{_\mu_\nu}\right)= 2 \kappa^2\left(\tensor{T }{_\mu_\nu}-\frac{1}{n} T \tensor{g}{_\mu_\nu}\right),\label{19}
\end{align}
 Choosing the gauge $g=1$, \cref{19} become the traceless EM of GR, \cite{Einstein1919,Ellis2011},
 \begin{equation}
 \tensor{R}{_\mu_\nu}-\frac{1}{n}R\tensor{g}{_\mu_\nu}=2 \kappa^2\left(\tensor{T }{_\mu_\nu}-\frac{1}{n} T \tensor{g}{_\mu_\nu}\right). \label{Em2} 
 \end{equation}
 Now in \cref{Em2}, there seems to be no CC. However, assuming the covariant conservation of the stress-energy tensor\footnote{The validity of \cref{34.5} is proven in \cref{sec:Sources} .} and using the Bianchi identities, 
 \begin{align}
 & \nabla_\m \left(T^{\m\n}-{1\over n}~T~g^{\m\n}\right)=-{1\over n}\nabla^\n T\rightarrow \label{34.5} \\
 &\nabla^\mu \left(\tensor{R}{_\mu_\nu}-\frac{1}{n}R\tensor{g}{_\mu_\nu}\right)=\frac{n-2}{2n}\nabla^\nu R=-\frac{2 \kappa^2}{n}\nabla^\nu T. 
 \end{align}
 In four dimensions, 
 \begin{equation}
 \nabla^\mu \left(2 \kappa^2T+R\right)=0\rightarrow 2 \kappa^2T+R=-\mathcal{C}.\label{10} 
\end{equation}
 Plugging in the constraint \cref{10} into the gauge fixed, four-dimensional EM, 
 \begin{equation}
 \tensor{R}{_\mu_\nu}-\frac{1}{2}R\tensor{g}{_\mu_\nu}+\mathcal{C}\tensor{g}{_\mu_\nu} =2 \kappa^2 \tensor{T }{_\mu_\nu}. \label{Em3} 
 \end{equation}
 In \cref{Em3}, the reader can appreciate the different context in which a \textit{Cosmological Constant term} appears in the EM as in contrast to GR. 
 Here it is an {integration constant} term, agnostic to the vacuum expectation value of fields.
 This discussion will be continued when considering its stability under radiative corrections in \cref{sec:quantization}.
 
 \section{Physical sources.}\label{sec:Sources}
 The previous discussion only considered the metric without any matter content. 
 The subtle issue of coupling matter fields to UG  is now considered in what follows. 
 Only scalar matter is considered for simplicity.
 \noindent The action of a scalar field minimally coupled to UG reads
 \begin{align}
 &S\equiv S_g+S_m=-\frac{1}{2\kappa^2}\int d^n\,x\,g^{\frac{1}{n}}\left(R\left[\tensor{g}{_\mu_\nu}\right]+\frac{(n-1)(n-2)}{4n^2}\frac{\nabla_\mu g\nabla^\mu g}{g^2}\right)+\nonumber\\
 &+\int d^n x \; \left[g^{1\over n}\left( {1\over 2}g^{\a\b}\pd_\a\phi\pd_\b\phi\right)-V(\phi)\right].
 \end{align}
 Defining, $2\kappa^2=M^{n-2}$, \cref{19} can be expressed as,
 \bea
 &&R_{\m\n}-{1\over n}~R~g_{\m\n}\equiv ~M^{2-n} \left(J^g_{\m\n}+J^m_{\m\n}\right)=\nonumber\\
 &&{(n-2)(2n-1)\over 4 n^2}\left({\nabla_\m g\nabla_\n g\over g^2}-{1\over n}{(\nabla g)^2\over g^2} g_{\m\n}\right)-{n-2\over 2n}\left({\nabla_\m\nabla_\n g \over g}-{1\over n}{\nabla^2 g\over g} g_{\m\n}\right)+\nonumber\\
 &&+{M^{2-n}\over 2}\left(\pd_\m \phi\pd_\n\phi-{1\over n}~g^{\a\b}\pd_\a\phi\pd_\b\phi~ g_{\m\n}\right).
 \eea 
Besides the matter part, $J^m_{\m\n}$, a gravitational piece $J^g_{\m\n}$ coming from the spacetime dependence of the metric determinant has been included as a source.
This is such that sources in the matter fields are defined according to, 
 \be
 g^{1\over n}~J_m^{\m\n}\equiv {\d S_m \over \d g_{\m\n}}.
 \ee
 Similarly, for the gravitational piece
 \be
 g^{1\over n}\left(R_{\m\n}-{1\over n} ~R~g_{\m\n}+M^{2-n}~J^g_{\m\n}\right)\equiv {\d S_g \over \d g_{\m\n}}.
 \ee
 \par In this work, the standard definition of the energy-momentum tensor in GR is used. For a scalar field with minimal coupling, this means
 \be
 T_{\m\n}\equiv \pd_\m \phi\pd_\n\phi-\left({1\over 2}g^{\a\b}\pd_\a\phi\pd_\b\phi-V(\phi)\right)g_{\m\n}.
 \ee
 Its trace is
 \be
 T\equiv g^{\a\b} T_{\a\b}=n V-{n-2\over 2}\left(\nabla\phi\right)^2,
 \ee
 so that the piece of the source that depends on the scalar field is precisely
 \be
 J^m_{\m\n}=T_{\m\n}-{1\over n}~T~g_{\m\n}=\pd_\m\phi\pd_\n\phi-{1\over n}~g^{\a\b}\pd_\a\phi\pd_\b\phi~g_{\m\n},
 \ee
 and does not include the potential energy.
 
 Let us revisit the covariant conservation equation used to prove \eqref{Em3},
 \be
 \nabla_\m \left(T^{\m\n}-{1\over n}~T~g^{\m\n}\right)=-{1\over n}\nabla^\n T.
 \ee
 This can be shown to hold in general precisely owing to the Ward identities of the area-preserving diffeomorphism plus Weyl symmetry.
 To be specific, the \textit{TDiff} Ward identities guarantee that there is a function $\Theta$ such that
 \be
 \nabla_\m \left(g^{{1\over n}-{1\over 2}} {\d S\over \d g _{\m\n}}\right)=\nabla^\n \Theta
 \ee
 The variation of the action under a variation of the scalar field reads \footnote{Here, the prime denotes differentiation with respect to the scalar field.} 
 \bea
 \d S_m&&=\int d^n x g^{1\over n} g^{\m\n}\pd_\n \d\phi-V^\prime \d\phi=-\int d^n x~\left(\pd_\n\left(g^{1\over n}~g^{\m\n}\pd_\m\phi\right)+V^\prime\right)\d\phi=\nonumber\\
 &&=\int d^n x g^{1\over n}\left(\nabla^2\phi-{n-2\over 2n}\nabla\phi.{\nabla g\over g}\right)+V^\prime (\phi).
 \eea
 Note that the EM of the scalar field have changed their usual form, and now they read,
 \be
 \nabla^2 \phi +g^{-{1\over n}} V^\prime(\phi) = {n-2\over 2n}{\nabla \phi.\nabla g\over g}.
 \ee
 It is then not difficult to check that
 \begin{align}
 \nabla_\m\left(g^{\frac{2-n}{2n}}\left(\nabla^\m\phi\nabla^\n\phi-{1\over n}(\nabla\phi)^2 g^{\m\n}\right)\right)&=g^{2-n\over 2n}\nabla^\n\left(V+{n-2\over 2n}\left(\nabla\phi\right)^2 \right)=\nonumber\\
 &=g^{2-n\over 2n}{1\over n}\nabla^\n T,
 \end{align}
 which modulo EM implies the desired result. Indeed, for any arbitrary constant value of $\l$,
 \be
 \nabla_\m \left(g^\l {\d S\over \d g_{\m\n}}\right)=\left(\nabla_\m g^\l\right){\d S\over \d g_{\m\n}}+g^\l\nabla_\m {\d S\over \d g_{\m\n}}.
 \ee
 Note that the additional term is multiplied by the EM. 
 \vspace{0.5cm}

 The remainder of this section will be devoted to showing that, although the only allowed source of the gravitational field in UG is just the traceless piece of the energy-momentum tensor
 \be
 J_{\m\n}\equiv T_{\m\n}-{1\over n} T g_{\m\n},
 \ee
 the {\em full Free Energy} produced by {\em arbitrary} sources (not only static ones) is {\em identical} to the one predicted by GR. This encompasses all weak field tests of gravitation.
 It is then convenient to restrict attention to the quadratic action corresponding to fluctuations around a given flat background coupled to arbitrary external sources, see \cite{Alvarez:2012px}. 
 
 \par To deal with an arbitrary background, a detailed analysis of the equations of motion will be performed to compare them to those of GR. The main conclusion here is that the UG ones are a subset of the ones corresponding to GR, with the source restricted to its traceless piece and with a vanishing CC.
 This refines a classical analysis \cite{Alvarez2005}. There it was argued that the Bianchi identities of UG imply a first integral of the equations of motion which, once used, made the unimodular theory fully equivalent to GR with an arbitrary CC. 
 Here a stronger result is proven, namely that this CC must vanish at the level of external sources. 
 \par
 Since they will play a crucial r\^ole in what follows, a slight detour is now taken to derive the Bianchi identities from the field theory viewpoint. 
 Start by defining a one form $\xi_1\equiv \xi_\m dx^\m$ as well as a two form $\Omega_2\equiv{1\over 2}\Omega_{\m\n}~dx^\m\wedge dx^\n$. The transversality condition now reads
 \be
 \xi_1=-2\d\Omega_2,
 \ee
 where the codifferential is the adjoint operator of the exterior derivative. Acting on two-forms
 \be
 \d\equiv *^{-1}~d~*,
 \ee
 so that its components obey 
 \be
 \left(\d\Omega_2\right)_\r=-{1\over 2}\nabla^\n \left(\Omega_2\right)_{\n\r}.
 \ee
 The number of independent components of a two-form is ${n\choose 2}$, but the codifferential is nilpotent $\d^2=0$, so one has to withdraw the three forms
 \be
 \Omega_2=\d \Omega_3,
 \ee
 and from them, one withdraws the four forms, etc. The final counting of independent gauge parameters is:
 \be
 {n\choose 2}-\left({n\choose 3}-\left({n\choose 4}-\cdots\right)\right)=n-1,
 \ee
 where the relationship $\sum_j \left(-1\right)^j~{n\choose j}=0$ has been used.
 
\noindent Taking into account that for antisymmetric tensors $\Omega^{\a\b}=-\Omega^{\b\a}\Rightarrow \nabla_\a\nabla_\b \Omega^{\a\b}\equiv 0$
 \bea
 &&0=\int d^n x ~\left(\nabla_\a \xi_\b+\nabla_\b \xi_\a\right)~{\d S\over \d g_{\a\b}}=\int d^n x ~\left(\nabla_\a \nabla^\rho\Omega_{\b\r}+\nabla_\b \nabla^\r\Omega_{\a\r}\right)~~{\d S\over \d g_{\a\b}}=\nonumber\\
 &&=-2 \int d^n x \sqrt{g}~\Omega^{\r\b}~ \nabla^\r\nabla_\a ~{1\over \sqrt{g}}~{\d S\over \d g_{\a\b}}.
 \eea
 Assuming that $\Omega^{\r\b}$ is arbitrary (they either are arbitrary or else vanishing), it follows that,
 \be
 \nabla^\r\nabla_\a ~{1\over \sqrt{g}}~{\d S\over \d g_{\a\b}}=\nabla^\b\nabla_\a ~{1\over \sqrt{g}}~{\d S\over \d g_{\a\r}}.
 \ee
 This is, for the vector
 \be
 \Theta^\b\equiv \nabla_\a ~{1\over \sqrt{g}}~{\d S\over \d g_{\a\b}},
 \ee
 one has the condition
 \be
 \nabla^\r\Theta^\b=\nabla^\b\Theta^\r,
 \ee
 which can be integrated as
 \be
 \Theta_\r=\nabla_\rho\Phi+\g_\r,
 \ee
 where $\g\equiv \g_\rho dx^\r$ is an harmonic form. The number of independent harmonic forms depends on the manifold's topology and is referred to as the first Betti number, $b_1(M)$, the dimension of the first cohomology group, $H^1(M)$.
 \par
 In case there are no harmonic forms in the spacetime manifold (which happens, in particular, if it is diffeomorphic to $\mathbb{R}^n$), this shows that the Bianchi identity is modified such that
 \be
 \nabla^\r\Theta^\b=\nabla^\b\Theta^\r=\nabla^\b\nabla^\rho\Phi.
 \ee
 i.e., still hold when integrated over the whole of spacetime with the \textit{Diff}  invariant measure.
 The Weyl invariance of the action means that
 \be
 0=\int d^n x~ w(x)~g_{\a\b}~{\d S\over \d g_{\a\b}},
 \ee
 which conveys the fact that, barring topological subtleties, the trace of the EM must be a total derivative
 \be
 g_{\a\b}~{\d S\over \d g_{\a\b}}=\pd_\r\Sigma^\r.
 \ee
 To get the Ward identities out of the action's symmetries, the parameters also need to be independent.
 The standard method starts with a change of variables in the path integral expressing the expectation value of a certain monomial of fields, $X\left[g_{\m\n},\psi_i\right]$, where $\psi_i$ is a generic representation of matter fields.
 \be
 Z \left\langle 0_+| X\left[g,\psi_i\right]|0_- \right\rangle\equiv e^{i W}\equiv \int {\cal D}g_{\m\n}~{\cal D}\psi~X[g_{\m\n},\psi_i]~e^{i S_{grav}[g]+i S_{matt}\left[g,\psi_i\right]}.
 \ee
 Namely $g_{\m\n}\rightarrow g_{\m\n}+\nabla_\m \xi_\n +\nabla_\n\xi_\m$, this leads easily to the Ward identity
 \begin{align}
 i \left\langle 0_+\left|{\d X\left[g,\psi_i\right] \over \d \Omega^{\m\n}(x)}\right|0_- \right \rangle=\nabla_\m \nabla^\a \left\langle 0_+\left|~X\left[g,\psi_i\right]~{ \d \tilde{S}\over \d g^{\a\n}(x)}\right|0_-\right \rangle-(\m \leftrightarrow \n).
 \end{align}

 In the particular case $X=1$, it states that the expectation value of the classical identity should vanish. There may be, in general, quantum corrections to the naive identities, either in the form of anomalies \cite{Jack} or even limit cycles \cite{Fortin}.
 
 \subsection{Free Energy with external sources in a flat background.}
 As mentioned above, to show the equivalence of UG and GR under all the weak field tests of gravitation, this section will make use of the Free Energy.
 The study of the Free Energy of UG neglecting self-interaction in the presence of external sources will show that it is fully equivalent to the GR one.
\par
First-order perturbations around a flat background correspond to,
 \begin{equation}
 g_{\m\n}\equiv \eta_{\m\n}+\kappa h_{\m\n}.
 \end{equation}
 Using the conventions of \cite{Alvarez2006}, in momentum space the kinetic part for UG reads\footnote{
 Compare this with the GR template, which to this order, corresponds to the Fierz-Pauli (FP) spin two theory. Its kinetic energy piece reads
 \begin{align}
 &8 K_{FP}^{\m\n\r\s}=k^2\left(\eta^{\m\r}\eta^{\n\s}+\eta^{\m\s}\eta^{\n\r}-2\eta^{\m\n}\eta^{\r\s}\right)\nonumber\\
 &-\left(k^\m k^\rho\eta^{\n\s}+k^\n k^\s \eta^{\m\r}+k^\m k^\s \eta^{\n\r}+k^\n k^\rho\eta^{\m\s}-2 k^\m k^\n \eta^{\r\s}-2 k^\rho k^\s \eta^{\m\n}\right).\nonumber
 \end{align}
 }
 \bea
 &&K^{U}_{\m\n\r\s}={1\over 8}k^2\left(\eta_{\m\r}\eta_{\n\s}+\eta_{\m\s}\eta_{\n\r}\right)-{1\over 8}\left(k_\n k_\s \eta_{\m\r}+\eta_{\m\s}k_\n k_\r+k_\n k_\r\eta_{\m\s}+k_\n k_\s\eta_{\m\r}\right)+\nonumber\\
 &&{1\over 2 n}\left(\eta_{\m\n}k_\rho k_\s+\eta_{\r\s}k_\m k_\n\right)-{n+2\over 4 n^2}k^2\eta_{\m\n}\eta_{\r\s}.
 \eea

 This can be expressed in terms of the Barnes-Rivers projectors; see Box 1.
 \\
 \begin{B0x}{1} \label{box1} 
 Starting with the longitudinal and transverse projectors
\begin{align}
 &\theta_{\a\b}\equiv\eta_{\a\b}-{k_\a k_\b\over k^2},\quad \omega_{\a\b}\equiv {k_\a k_\b\over k^2}. \tag{B.1}
\end{align}
 They obey
\begin{align}
 &\theta+\omega\equiv \theta_\m^\n+\omega_\m^\n=\d_\m^\n\equiv 1,\nonumber\\
 &\theta^2\equiv \theta_\a^\b\theta_\b^\g=\theta_\a^\g\equiv \theta,\nonumber\\
 &\omega^2\equiv \omega_\a^\b \omega_\b^\g=\omega_\a^\g\equiv \omega,\nonumber\\
 & tr~\theta=n-1,\nonumber\\
 &tr~\omega=1.\tag{B.2}
\end{align}
 The four-indices projectors are
\begin{align}
 &P_2\equiv {1\over 2}\left(\theta_{\m\r}\theta_{\n\s}+\theta_{\m\s}\theta_{\n\r}\right)-{1\over n-1}\theta_{\m\n} \theta_{\r\s},\nonumber\\
 &P_1\equiv{1\over 2}\left(\theta_{\m\r}\omega_{\n\s}+\theta_{\m\s}\omega_{\n\r}+\theta_{\n\r}\omega_{\m\s}+\theta_{\n\s}\omega_{\m\r}\right),\nonumber\\
 &P_0^s\equiv {1\over n-1}\theta_{\m\n}\theta_{\r\s},\nonumber\\
 &P_0^w\equiv \omega_{\m\n}\omega_{\r\s},\nonumber\\
 &P_0^{sw}\equiv{1\over \sqrt{n-1}}\theta_{\m\n}\omega_{\r\s},\nonumber
\end{align}
\begin{align}
 &P_0^{ws}\equiv{1\over \sqrt{n-1}}\omega_{\m\n}\theta_{\r\s}.\tag{B.3}
\end{align}
 They obey
 \begin{align}
 &P_i^a P_j^b=\d_{ij}\d^{ab} P_i^b,\nonumber\\
 &P_i^a P_j^{bc}=\d_{ij}\d^{ab}P_j^{ac},\nonumber\\
 &P_i^{ab} P_j^c=\d_{ij}\d^{bc} P_j^{ac},\nonumber\\
 &P_i^{ab} P_j^{cd}=\d_{ij}\d^{bc}\d^{ad} P_j^a. \label{72} \tag{B.4}
 \end{align}
 as well as
\begin{align}
 &tr~P_2\equiv \eta^{\m\n} (P_2)_{\m\n\r\s}=0,\nonumber\\
 &tr~P_0^s=\theta_{\r\s},\nonumber\\
 &tr~P_0^w=\omega_{\r\s},\nonumber\\
 &tr~P_1=0,\nonumber\\
 &tr~P_0^{sw}=\sqrt{n-1}~\omega_{\r\s},\nonumber\\
 &tr~P_0^{ws}={1\over \sqrt{n-1}}~\theta_{\r\s},\nonumber\\
 &P_2+P_1+P_0^w+P_0^s={1\over 2}\left(\d_\m^\n \d_\r^\s+\d_\m^\s \d_\r^\n\right). \tag{B.5}
\end{align}
 
 Any symmetric operator can be written as
 \be
 K= a_2 P_2 + a_1 P_1 + a_w P_0^w + a_s P_0^s + a_\times P_0^\times,\tag{B.6}
 \ee
 (where $P_0^\times\equiv P_0^{ws}+P_0^{sw}$).
 Then
 \be
 K^{-1}={1\over a_2}P_2+{1\over a_1} P_1 +{a_s\over a_s a_w - a_\times^2}P_0^w+{a_w\over a_s a_w - a_\times^2}P_0^s-{a_\times\over a_s a_w - a_\times^2}P_0^\times. \tag{B.7}
 \ee
 
Define the trace-free projector
 \be
 \left(P_{tr}\right)_{\r\s}\,^{\l\d}\equiv {1\over 2}\left(\d_\r^\l \d_\s^\d+\d_\r^\d \d_\s^\l\right)-{1\over n}\eta_{\r\s}\eta^{\l\d}. \tag{B.8}
 \ee
 
Then, the following relations hold:
\begin{align}
 &\left(P_2\right)_{\m\n}\,^{\r\s}\left(P_{tr}\right)_{\r\s}\,^{\l\d}=P_2,\nonumber\\
 &P_0^s P_{tr}=P_0^s-{n-1\over n}P_0^s-{\sqrt{n-1}\over n}P_0^{sw},\nonumber\\
 &P_0^w P_{tr}=P_0^w-{\sqrt{n-1}\over n}P_0^{ws}-{1\over n}P_0^w,,\nonumber\\
 &P_1 P_{tr}=P_1,\nonumber\\
 &P_0^{sw} P_{tr}=P_0^{sw}-{\sqrt{n-1}\over n}P_0^{ws}-{1\over n}P_0^w,\nonumber\\
 &P_0^{ws} P_{tr}=P_0^{ws}-{\sqrt{n-1}\over n} P_0^{sw}-{n-1\over n} P_0^s.\tag{B.9}
\end{align}
 \end{B0x}
 \vspace{0.4cm}
 From the properties presented in Box 1,
 \bea
 &&K^{U}={k^2\over 8}\left(2 P_2+ 2 P_0^s+ 2 P_1+ 2 P_0^w\right)-{k^2\over 8}\left(2 P_1 + 4 P_0^w\right)+\nonumber\\
 &&{k^2\over 2 n}\left(\sqrt{n-1}P_0^\times+ 2 P_0^w\right)-{n+2\over 4 n^2}k^2 \left(\left(n-1\right)P_0^s+\sqrt{n-1}P_0^\times + P_0^w\right)=\nonumber\\
 &&k^2\bigg\{{1\over 4}P_2-{n-2\over 4 n^2} P_0^s-{ n^2-3 n +2\over 4 n^2}P_0^w+{n-2\over 4 n^2}\sqrt{n-1}P_0^\times\bigg\}. 
 \eea
 Then, it is plain that
 \bea
 &&K^{WT}_{\m\n\r\s}\eta^{\r\s}=0,\nonumber\\
 && \xi.k=0~\Rightarrow~K_{\m\n\r\s}\xi^\rho k^\s=0.
 \eea
 To gauge fix the UG action, consider adding a gauge fixing term:
 \be
 \mathcal{L}_{gf}=h_{\m\n}~K_{gf}^{\m\n\r\s}~h_{\r\s},
 \ee
 where using \cref{72},
 \be
 K_{gf}^{\m\n\r\s}={k^6\over 4 \Lambda^4}P_1.
 \ee
 This corresponds in position space to a gauge fixing
 \be
 \mathcal{L}_{gf}={1\over 2 \Lambda^4} F_\a^2\equiv {1\over 2 \Lambda^4}\left(\pd_\a\pd^\m\pd^\n h_{\m\n}-\Box \pd^\m h_{\a\m}\right)^2,
 \ee
 where $\Lambda$ is an arbitrary mass scale. Let us remark that this gauge choice is admissible because it can be reached uniquely through area-preserving diffeomorphism
 \bea
 &&\pd_\a F^\a=0,\nonumber\\
 &&\d F_\a=-\Box^2 \xi_\a.
 \eea
The ghost system associated with it gets complicated because the gauge parameters are not independent\footnote{A complete analysis can be found in \cite{AlvarezV}.}, more will be said regarding this fact when dealing with one-loop UG in \cref{sec:quantization}. However, this issue is irrelevant for the purposes at hand, which are purely tree-level.
 
At the end of the day, 
 \bea
 &&4 K_{tot}^{U}=k^2 P_2+\left((n-1) m^2-{n-2\over n^2}k^2\right) P_0^s+\left(m^2-{ n^2-3 n +2\over n^2}k^2\right) P_0^w+\nonumber\\
 &&\left(m^2+{n-2\over n^2}k^2\right)\sqrt{n-1}P_0^\times+
 {k^6\over M^4}P_1.
 \eea
Therefore the euclidean propagator is then given, for the gauge choice above, by
 \bea
 &&k^2 \Delta= P_2+{M^4\over k^4} P_1-{1\over (n-2) m^2}\bigg\{\left(m^2-{n^2-3n+2\over n^2}k^2\right) P_0^s+\nonumber\\
 &&\left((n-1)m^2 -{n-2\over n^2}k^2\right) P_0^w-\left(m^2+{n-2\over n^2}k^2\right)\sqrt{n-1} P_0^\times\bigg\}.
 \eea
 Any coupling of the gravitational fluctuation to an external source $S_{int}=\int d^n x~J_{\m\n} h^{\m\n}$ has to comply with
 \be
 0=\d S_{int}=\int d^n x~J_{\m\n}\left(\pd_\m \pd_\r\Omega^\r_\m+\pd_\n \pd_\r\Omega^\r_\n+ \omega(x) h_{\m\n}\right),
 \ee
 which requires the sources to obey both $\eta_{\m\n} J^{\m\n}=0$ and $\pd_\m J^{\m\n}=\pd^\n T$. 
 This means that it should be related to some conserved symmetric tensor\footnote{ A priori this could be different from the usual conserved energy-momentum tensor although it will be proven to be the same.} by 
 \be
 J_{\m\n}\equiv T_{\m\n}-{1\over n} T \eta_{\m\n}.
 \ee
 The Free Energy (or effective action) after the gaussian functional integration reads
 \bea
 W[J]&&\equiv {1\over 2}\int d^n x d^n y J^*_{\m\n}(x)~\Delta^{\m\n\r\s}(x,y)~J_{\r\s}(y)=\nonumber\\
 &&={1\over 2}(2\pi)^{2n} \int d^n k 
 J_{\m\n}^*(k)~\Delta^{\m\n\r\s}(k)~J_{\r\s}(k),
 \eea
 and using the easily proven identities
 \bea
 &&\left(P_1 J\right)_{\m\n}=0,\nonumber\\
 &&tr\,J P_1 J=0,\nonumber\\
 &&(P_2)_{\m\n\r\s} J^{\r\s}=T_{\m\n}-{1\over n-1}\theta_{\m\n} T,\nonumber\\
 &&tr\, \left(J P_2 J\right)= |T_{\m\n}|^2-{1\over n-1}~|T|^2,
 \eea
 it conveys the fact that
 \be
 W[J]={1\over 2}(2\pi)^{2n} \int d^n k \bigg\{
 J_{\m\n}^*(k)~{1\over k^2}~P_2^{\m\n\r\s}(k)~J_{\r\s}(k)+C(k) |T(k)|^2\bigg\},
 \ee
 where
 \be
 C(k)=-{1\over (n-1)(n-2)k^2}.
 \ee
 This yields the Free Energy
 \bea\label{free}
 &&W[T]={1\over 2}(2\pi)^{2n} \int d^n k 
 {1\over k^2}\left(|J_{\m\n}|^2-{2\over n(n-2)}|T(k)|^2\right)=\nonumber\\
 && ={1\over 2}(2\pi)^{2n} \int d^n k 
 {1\over k^2}\left(|T_{\m\n}|^2-{1\over n-2}|T(k)|^2\right).
 \eea
 This Free Energy (when expressed in the second form) is {\em exactly} the same as the prediction of GR, which implies that the low energy physics, and so the low energy empirical tests of UG, are the same as in GR and convey the same results. Only in the non-linear regime could some differences between both theories be found.
 \\
 Consider as a particular example the computation of the Newtonian potential.
 In UG, the graviton field $h_{\m\n}$ couples to the traceless part, $\hat{T}^{\m\n}$, of the Energy-momentum tensor {\it \`a la} Rosenfeld or, what is the same, the traceless part of the graviton field, $\hat{h}_{\m\n}$, couples to
 the the Energy-momentum tensor defined {\it \`a la} Rosenfeld:
 \be -\frac{\kappa}{2}\int d^{4}x\;h_{\m\n}\,\hat{T}^{\m\n}=-\frac{\kappa}{2}\int d^{4}x\;\hat{h}_{\m\n}T^{\m\n},
 \label{mattercoupling}
 \ee
 where
 \be \hat{T}^{\m\n}=T^{\m\n}-\frac{1}{4}\,T\,\eta^{\m\n}\quad\text{and}\quad \hat{h}_{\m\n}=h_{\m\n}-\frac{1}{4}\,h\,\eta_{\m\n}.
 \ee
The Newtonian potential can then be obtained \cite{Donoghue:1994dn} from the tree-level one-graviton exchange, with transfer momentum $k_\m$, between two very massive scalar particles by taking the static limit: $k_\m=(0,\vec{k})_\m$. Let
 $\mathcal{A}_{12}$ denote the amplitude for the one-graviton exchange between two scalar particles with masses $M_1$ and $M_2$, respectively. 
 In UG --see  \cref{mattercoupling}-- one has
 \be
 \mathcal{A}_{12}=-i\frac{\kappa^2}{4}T_{\m\n}^{1}(p_1,p'_1)<\hat{h}^{\m\n}(k) \hat{h}^{\r\s}(-k)> T_{\r\s}^{2}(p_2,p'_2),
 \ee
 where $k=p_1-p'_1=p'_2-p_2$. In the previous equation, $<\hat{h}^{\m\n}(k) \hat{h}^{\r\s}(-k)>$ denotes the free two-point function of the traceless graviton field, and $T_{\m\n}^{i}(p_i, p'_i)$, $i=1,2$, denote the lowest order contribution to the
 on-shell matrix elements of the energy-momentum tensor between (on-shell) states with momentum $p_i$ and $ p'_i$, $i=1,2$, respectively:
 \be T_{\m\n}^{i}(p_i,p'_i)=p_{i\,\m}p'_{i\,\n}+p_{i\,\n}p'_{i\,\m}+\frac{1}{2}\,k^2\eta_{\mu\n}.
 \ee
 Now, for very massive particles and for $k_\mu=(0,\vec k)$, 
 \be \frac{1}{2M_i}T_{\m\n}^{i}(p_i,p'_i)= M_{i}\,\eta^{\m 0}\eta^{\n 0},\quad i=1,2
 \ee
 so that, in the static limit, 
 \be
 \frac{1}{2M_1\,2M_2}\mathcal{A}_{12}=-i\frac{\kappa^2}{4}\,m_1\,m_2\,<\hat{h}^{00}(k) \hat{h}^{00}(-k)>,
 \label{nonrelati}
 \ee
 with $k_\m=(0,\vec{k})_\m$. It is the RHS of the previous equation which must be equal to the Newtonian potential in Fourier space $V_{Nw}(\vec{k})$, where
 \be
 V_{Nw}(\vec{k})=-\frac{\kappa^2}{8}\, \frac{M_1 M_2}{\vec{k}^2}.
 \label{Newtpotential}
 \ee
 
 Assuming that in UG the free graviton two-point function, $<h_{\m\n}(k)h_{\r\s}(-k)>$, correspond to,
 \begin{align}
 \langle h_{\m\n}(k)h_{\r\s}(-k)\rangle&=\frac{i}{2k^2}\left(\eta_{\m\s}\eta_{\n\r}+\eta_{\m\r}\eta_{\n\s}-\eta_{\m\n}\eta_{\r\s}\right)-\frac{a(k^2)}{2k^2}\eta_{\m\n}\eta_{\r\s}+\nonumber\\
 &+\frac{b(k^2)}{(k^2)^2}\left(k_\rho k_\s \eta_{\m\n}+k_\m k_\n \eta_{\r\s}\right)+\frac{c(k^2)}{(k^2)^3}\,k_{\m}k_{\n}k_{\r}k_{\s},
 \label{propansatz}
 \end{align}
 where $a(k^2)$, $b(k^2)$ and $c(k^2)$ are arbitrary real functions. The ansatz in \cref{propansatz} is the most general expression consistent with Lorentz covariance, boson symmetry, the symmetry of $h_{\m\n}$ and that when one replaces in the free two-point function, the tensor
 \be
 \frac{1}{2}\left(\eta_{\m\s}\eta_{\n\r}+\eta_{\m\r}\eta_{\n\s}-\eta_{\m\n}\eta_{\r\s}\right).
 \ee
 with the following sum over polarizations,
 \be
 \sum_{\lambda=\pm 2}\e^{(\lambda)}_{\m\n}\e^{(-\lambda)}_{\r\s}.
 \ee
Only a simple pole factor $1/k^2$ multiplies this sum, as it befits the unitarity and the fact that the theory's classical action is quadratic on the derivatives.
 
 From \cref{propansatz}, one obtains after a little algebra
 \begin{align}
 \langle \hat{h}_{\m\n}(k)\hat{h}_{\r\s}(-k)\rangle&= \frac{i}{2k^2}\left(\eta_{\m\s}\eta_{\n\r}+\eta_{\m\r}\eta_{\n\s}+\big(-\frac{1}{2}+\frac{c(k^2)}{8}\big)\eta_{\m\n}\eta_{\r\s}\right)+\nonumber\\
 &+\frac{c(k^2)}{4(k^2)^2}\left(k_\rho k_\s \eta_{\m\n}+k_\m k_\n \eta_{\r\s}\right)+\frac{c(k^2)}{(k^2)^3}\,k_{\m}k_{\n}k_{\r}k_{\s}.
 \end{align}
 Substituting the previous result in \cref{nonrelati} --recall that $k_\m=(0,\vec{k})_\m$,
 \be
 -i\frac{\kappa^2}{4}\,m_1\,m_2\,\langle\hat{h}^{00}(k) \hat{h}^{00}(-k)\rangle =-\frac{\kappa^2}{8}\,M_1\,M_2\,\big(\frac{3}{2}+\frac{c(-\vec{k}^2)}{8}\big)\,\frac{1}{\vec{k}^2}.
 \ee
 This expression will match the Newtonian potential in (\ref{Newtpotential}) if, and only if, $c(-\vec{k}^2)=-4$, which, by Lorentz invariance leads to
 \be
 c(k^2)=-4,
 \ee
 regardless of the value of $k_\mu$. 
 
 In summary, a triple pole in the $k_{\m}k_{\n}k_{\r}k_{\s}$ contribution to the two-point function in \cref{propansatz} is needed to get the Newtonian potential right. This is what happens when one
 works out the propagator of UG by using the BRST technique explained in \cite{AlvarezV,Alvarez2015} and discussed in \cref{sec:quantization} below. Notice that when $n=4$, the propagator in \cref{propagatorug} yields the Newtonian potential since the coefficient multiplying
 the contribution
 \be
 \dfrac{k_{\m}k_{\n}k_{\r}k_{\s}}{k^6},
 \ee
 is $-4$, at $n=4$.

 \subsection{External sources with a general background.}\label{sec:FE2} 
 In the presence of vacuum energy, Minkowski space is not a solution. Hence, it is not an appropriate background.
 This section considers an arbitrary background $\bg_{\m\n}$ instead. Therefore, a closed form for the Free Energy will not be available in this case. 
 The analysis here will rely instead on a detailed analysis of the field equations endowed with arbitrary sources. The conclusion will again be that the unimodular EM are equivalent to the GR ones with vanishing CC. 
 Expanding the Lagrangian in \cref{UGact1}  up to the second order in the metric perturbation around an arbitrary background,
 \begin{equation}
 g_{\mu\nu}=\bg_{\m\n}+\kappa h_{\m\n},
 \end{equation}
 reads;
 \bea
 &&\mathcal{L}_U\equiv g^{1\over n} R+{(n-1)(n-2)\over 4 n^2}{(\nabla g)^2\over g^2}=\nonumber\\
 &&\bg^{1\over n}\bigg[\bR+\kappa\left({h\over n}\bR-h^{\a\b} \bR_{\a\b}-\bn^2 h + \bn_\a \bn_\b h^{\a\b}\right)+{\kappa^2\over 4}\bigg\{-2\bn_\b\left(h^{\a\r}\bn^\b h_{\a\r}\right)+\nonumber\\
 &&+2\bn_\a\left( 2 h^{\a\r}\bn^\b h_{\r\b}-h^{\a\r}\bn_\rho h\right)-2 h^{\b\n}\left(\bn^\r\bn_\n h_{\r\b}+\bn^\r\bn_\b h_{\r\n}-\bn^2 h_{\n\b}-\bn_\b\bn_\n h\right)-\nonumber\\
 &&-2 \bn_\s h\bn^\s h+2 \bn_\s h\bn_\b h^{\s\b}- 2 \bn_\s h_{\a\n}\bn^\n h^{\a\s}+ \bn_\s h_{\a\n} \bn^\s h^{\a\n}+ 4 h^\b_\a h^{\n\a} \bR_{\n\b}-\nonumber\\
 &&{4\over n} h \left(h^{\a\b}\bR_{\a\b}+\bn^2 h-\bn_\a\bn_\b h^{\a\b}\right)+{2\over n}\left({h^2\over n}-h^{\a\b}h_{\a\b}\right)\bR \bigg\}\bigg]+\nonumber\\
 &&+{(n-1)(n-2)\over 4 n^2}\bigg\{\kappa^2\bigg(-2h^{\m\n}\bn_\m h{\bn_\n\bg\over\bg}+(\bn h)^2-2(h\bn_\m h+h^{\a\b}\bn_\m h_{\a\b}){\bn^\m\bg\over\bg}+\nonumber\\
 &&+h^{\m\a}h^\n_\a {\bn_\m\bg\over\bg}{\bn_\n\bg\over\bg}\bigg)+{(\bn\bg)^2\over \bg^2}+\kappa\left(2\bn_\m h {\bn^\m\bg\over \bg}-h^{\m\n}{\bn_\m\bg\over \bg}{\bn_\n\bg\over\bg}\right)\bigg\}.
 \eea
 
 The only way the linear term can vanish is by restricting either the allowed fluctuations or else the allowed backgrounds through 
 \be
 {h\over n}\bR-h^{\a\b} \bR_{\a\b}-\bn^2 h + \bn_\a \bn_\b h^{\a\b}+2\bn_\m h {\bn^\m\bg\over \bg}-h^{\m\n}{\bn_\m\bg\over \bg}{\bn_\n\bg\over\bg}=0. \label{114} 
 \ee
For maximally symmetric backgrounds, in which,
\begin{equation}
  \bR_{\m\n}=-{2\l\over n-2}\bg_{\m\n}, \label{maxsim} 
\end{equation}
\cref{114} reads
 \be
 \bn^2 h - \bn_\a \bn_\b h^{\a\b}+2\bn_\m h {\bn^\m\bg\over \bg}-h^{\m\n}{\bn_\m\bg\over \bg}{\bn_\n\bg\over\bg}=0.
 \ee
 A simple solution is restricting the background to be unimodular, $\bg=1$. 
 Then the offending terms either vanish or else behave as total derivatives.
Once a unimodular background is chosen, the linear term corresponds to the EM for the background field
 \begin{align}
 h^{\mu\nu}\left( \bar{R}_{\mu\nu}-\frac{1}{n}\bar{R}\bar{g}_{\mu\nu}\right)=0.
 \end{align}
 The analysis of the EM of both theories around arbitrary backgrounds $\bar{g}_{\m\n}$ and $\hat{g}_{\m\n}$ can now be done.
 The lagrangians for UG and GR with a CC\footnote{
 In the full non-linear theory, the CC is included in an arbitrary energy-momentum tensor. In the linear approximation, this is not the case.
 }, both expanded up to second order in linear perturbations, are
 \begin{align}
 \mathcal{L}_{U}&=\frac{n+2}{4n^{2}}\bar{\nabla}^{\mu}h\bar{\nabla}_{\mu}h-\frac{1}{n}\bar{\nabla}_{\mu}h\bar{\nabla}^{\rho}h^{\mu}_{\rho}+\frac{1}{2}\bar{\nabla}_{\mu}h^{\mu\rho}\bar{\nabla}_{\nu}h^{\nu}_{\rho}-\frac{1}{4}\bar{\nabla}_{\mu}h^{\nu\rho}\bar{\nabla}^{\mu}h_{\nu\rho}-\nonumber\\
 &-\bar{R}_{\nu\beta}h^{\beta}_{\alpha}h^{\nu\alpha}+\frac{1}{n}h\bar{R}_{\alpha\beta}h^{\alpha\beta}-\frac{\bar{R}}{2}\left( \frac{h^{2}}{n^{2}}-\frac{1}{n}h^{\alpha\beta}h_{\alpha\beta}\right),\label{118} 
 \end{align}
 and,
 \begin{align}
 \mathcal{L}_{GR\lambda}&=\frac{1}{4}\hat{\nabla}^{\mu}h\hat{\nabla}_{\mu}h-\frac{1}{2}\hat{\nabla}_{\mu}h\hat{\nabla}^{\rho}h^{\mu}_{\rho}+\frac{1}{2}\hat{\nabla}_{\mu}h^{\mu\rho}\hat{\nabla}_{\nu}h^{\nu}_{\rho}-\frac{1}{4}\hat{\nabla}_{\mu}h^{\nu\rho}\hat{\nabla}^{\mu}h_{\nu\rho}-\nonumber\\
 &-\hat{R}_{\nu\beta}h^{\beta}_{\alpha}h^{\nu\alpha}+\frac{1}{2}h\hat{R}_{\alpha\beta}h^{\alpha\beta}-\frac{\hat{R}+2\lambda}{2}\left( \frac{h^{2}}{4}-\frac{1}{2}h^{\alpha\beta}h_{\alpha\beta}\right).\label{119} 
 \end{align}
 
 \vspace{0.1cm}
 Assuming both backgrounds to correspond to maximally symmetric spaces \cref{maxsim}, \cref{118,119} reduce to 
 \begin{align}
 \mathcal{L}_{U}&=\frac{n+2}{4n^{2}}\bar{\nabla}^{\mu}h\bar{\nabla}_{\mu}h-\frac{1}{n}\bar{\nabla}_{\mu}h\bar{\nabla}^{\rho}h^{\mu}_{\rho}+\frac{1}{2}\bar{\nabla}_{\mu}h^{\mu\rho}\bar{\nabla}_{\nu}h^{\nu}_{\rho}-\frac{1}{4}\bar{\nabla}_{\mu}h^{\nu\rho}\bar{\nabla}^{\mu}h_{\nu\rho},\\
 \mathcal{L}_{GR\lambda}&=\frac{1}{4}\hat{\nabla}^{\mu}h\hat{\nabla}_{\mu}h-\frac{1}{2}\hat{\nabla}_{\mu}h\hat{\nabla}^{\rho}h^{\mu}_{\rho}+\frac{1}{2}\hat{\nabla}_{\mu}h^{\mu\rho}\hat{\nabla}_{\nu}h^{\nu}_{\rho} -\frac{1}{4}\hat{\nabla}_{\mu}h^{\nu\rho}\hat{\nabla}^{\mu}h_{\nu\rho}-\nonumber\\
 &-\frac{\lambda}{2}\left( \frac{h^{2}}{2}-h_{\alpha\beta}h^{\alpha\beta}\right).
 \end{align}
 Sources for both theories can be introduced in the usual way by a linear coupling. 
 In the case of GR, the source is just the usual symmetric energy-momentum tensor $T_{\mu\nu}$ while UG, as discussed above, couples to the traceless source $J_{\mu\nu}$ which obeys $\nabla_\m J^{\m\n}=\nabla^\n T$. 
 
 \vspace{0.4cm}
 The EM for UG, dubbed EMU, read
 \bea
 &&EMU\equiv \frac{n+2}{2n^{2}}\bar{g}_{\mu\nu}\bar{\nabla}^{2}h-\frac{1}{2}\bar{\nabla}^{2}h_{\mu\nu}-\frac{1}{n}\bar{\nabla}_{\alpha}\bar{\nabla}_{\beta}h^{\alpha\beta}\bar{g}_{\mu\nu}-\nonumber\\
 && \frac{1}{n}\bar{\nabla}_{\mu}\bar{\nabla}_{\nu}h + \frac{1}{2}\bar{\nabla}_{\mu}\bar{\nabla}_{\alpha}h^{\alpha}_{\nu}+\frac{1}{2}\bar{\nabla}_{\nu}\bar{\nabla}_{\alpha}h^{\alpha}_{\mu}=J_{\mu\nu},
 \eea
 whereas the EM of general relativity, EMGR, are
 \bea
 &&EMGR\equiv \frac{1}{2}\hat{\nabla}^{2}h\hat{g}_{\mu\nu}-\frac{1}{2}\hat{\nabla}^{2}h_{\mu\nu}-\frac{1}{2}\hat{\nabla}_{\alpha}\hat{\nabla}_{\beta}h^{\alpha\beta}\hat{g}_{\mu\nu}-\frac{1}{2}\hat{\nabla}_{\mu}\hat{\nabla}_{\nu}h+\nonumber\\
 && \frac{1}{2}\hat{\nabla}_{\mu}\hat{\nabla}_{\alpha}h^{\alpha}_{\nu}+\frac{1}{2}\hat{\nabla}_{\nu}\hat{\nabla}_{\alpha}h^{\alpha}_{\mu}= \lambda \left(\frac{h}{2}\hat{g}_{\mu\nu}-h_{\mu\nu}\right)+T_{\mu\nu}.
 \eea
 At this point, the result advertised in the introduction on the equivalence of the unimodular theory with GR with an undetermined CC should be remembered. 
 Fluctuations around a flat background were already analyzed in the previous section, and full equivalence with GR with vanishing CC was found. 
 To ensure that this result is not an artifact of the flat background, it is worth repeating the analysis in this more general setting.
 \par
 The first integrals from the EM can be derived to make things easy. The first one, $I_{GR}$, by taking the trace of the EMGR:
 \be
 I_{GR}\equiv \hat{\nabla}^{2}h-\hat{\nabla}_{\alpha}\hat{\nabla}_{\beta}h^{\alpha\beta}-\lambda h=\frac{2}{n-2}T,\\
 \ee
 whereas the second one stems from taking the covariant divergence of the EMU
 \bea
 &&I_U\equiv \frac{2-n}{2n}\bar{\nabla}_{\nu}\left( \bar{\nabla}_{\alpha}\bar{\nabla}_{\beta}h^{\alpha\beta}-\frac{1}{n}\bar{\nabla}^{2}h+\frac{1}{n}T\right)=0 \nonumber\\
 &&\Rightarrow \bar{\nabla}_{\alpha}\bar{\nabla}_{\beta}h^{\alpha\beta}-\frac{1}{n}\bar{\nabla}^{2}h+\frac{1}{n}T=\Gamma,
 \eea
 where $\Gamma$ is an arbitrary constant.
 Assume that the background metric is the same for both theories\footnote{This is a reasonable ansatz to check equivalence.}, $\bar{g}_{\mu\nu}=\hat{g}_{\mu\nu}$. 
 After that, perform a field redefinition of the form $h_{\mu\nu}=H_{\mu\nu}+a H\bar{g}_{\mu\nu}$ that would map one theory into the other, with the possible addition of terms proportional to the first integrals which are zero by the use of the EM. 
 This is equivalent to a search for constants $a$, $C_{1}$ $C_{2}$ and $\Gamma$ such that
 \begin{align}
  EMGR&\left(H_{\m\n}+ a H \bg_{\m\n}\right)+ C_2 I_{GR}\left(H_{\m\n}+ a H \bg_{\m\n}\right)=\nonumber\\
  &= EMU\left(H_{\m\n}\right)+C_1 I_U\left(H_{\m\n}\right),
 \end{align}
 and
 \bea
 &&\bar{\nabla}^{2}H \bar{g}_{\mu\nu}\left( \frac{n+2}{2n^{2}}-\frac{C_{1}}{n}-\frac{1}{2}-a\left( \frac{n}{2}-1\right)-C_{2}(1+na-a) \right)+\nonumber\\
 &&+\bar{\nabla}_{\alpha}\bar{\nabla}_{\beta}H^{\alpha\beta}\bar{g}_{\mu\nu}\left( C_{1}+C_{2}+\frac{1}{2}-\frac{1}{n}\right)+\bar{\nabla}_{\mu}\bar{\nabla}_{\nu}H\left( \frac{1}{2}-\frac{1}{n}-a\left( 1-\frac{n}{2}\right)\right)+\nonumber\\
 &&+H\bar{g}_{\mu\nu}\lambda\left(\frac{1}{2}+a\left( \frac{n}{2}-1\right)+C_{2}(1+na)\right)+\bar{g}_{\mu\nu}\left( T\left( \frac{2C_{2}}{n-2}+\frac{C_{1}+1}{n}\right)-C_1 \Gamma\right)+\nonumber\\
 &&+T_{\m\n}-J_{\m\n}-
 \lambda H_{\mu\nu}=0.
 \eea
 
 The equations obtained by demanding every factor to be zero are only compatible if the CC $\lambda$ vanishes. In that case, the solution of the system is simply
 \bea
 &&a=-\frac{1}{n},\nonumber\\
 && C_{1}+C_{2} = \frac{2-n}{2n},\nonumber\\
 && \Gamma = \left( \frac{n^{2}(2C_{2}-1)+n(4C_{2}+4)-4}{2n^{2}(n-2)}\right) T.
 \eea
 To summarize,
 \begin{align}
 EMGR&\left(H_{\m\n}-{1\over n} H \bg_{\m\n}\right) -{n-2\over 2n}I_{GR}\left(H_{\m\n}-{1\over n} H \bg_{\m\n}\right)=\nonumber\\
 &=EMU\left(H_{\m\n}\right)\left.\right|_{J_{\m\n}=T_{\m\n}-{1\over n} T \bg_{\m\n}}.
 \end{align}
 The physical meaning of what has been proved is that the unimodular EMU are a consequence of EMGR when $\l=0$ {\em only}; it is the subsector corresponding to
 \be
 h^{GR}_{\m\n}=h^U_{\m\n}-{1\over n} h^U g_{\m\n}.
 \ee
 It is worth remarking (again) that this is {\em not} a field redefinition; it is a truncation of GR such that $h^{GR}=0$. There is no way to build the inverse map from EMU to EMGR. Given the fact that
 \be
 h=0,
 \ee
 is a (partial) algebraic gauge fixing (which does not need ghosts); this shows that, at the level of the EM, the unimodular theory is a truncation of GR with vanishing CC and with the source reduced to the traceless part of the GR source. It is perhaps worth remarking that this does not necessarily follow from the fact that the Lagrangian is so obtained (gauge conditions can only be used {\em after} the EM are derived).
 \section{No Birkhoff theorem in UG.}\label{sec:Birkhoff}
 It would seem that Birkhoff's theorem trivially holds in UG. After all, it is claimed, \cite{Alvarez2005}, that UG is (classically at least) equivalent to GR with a CC unrelated to vacuum energy. So this would seem to close the argument. 
 \par
 Another urban myth is that Birkhoff's theorem is related to the fact that GR only propagates spin two without any spin zero impurities. Given that UG also propagates spin two only, it should follow that Birkhoff's theorem should also hold for UG.
 \par
 Things are not so simple; recall the vacuum EM for UG derived in \cref{sec:Non-Linear Field Theory};
 \begin{align}
  &\tensor{R}{_\mu_\nu}-\frac{1}{n}R\tensor{g}{_\mu_\nu}+\frac{(2-n)(2n-1)}{4n^2}\left(\frac{\nabla_\mu g\nabla_\nu g}{g^2}-\frac{1}{n}\frac{(\nabla g)^2}{g^2}\tensor{g}{_\mu_\nu}\right)+\nonumber\\
  &+\frac{n-2}{2n}\left(\frac{\nabla_\mu\nabla_\nu g}{g}-\frac{1}{n}\frac{\nabla^2g}{g}\tensor{g}{_\mu_\nu}\right)=0. \label{emugbis}
 \end{align}
 In the Weyl gauge $g=1$, they reduce to 
 \begin{equation}
 R_{\m\n}={1\over n} \,R\, g_{\m\n}.\label{emugtrbis}
\end{equation}
 The way to solve \cref{emugbis} is first to solve \cref{emugtrbis} for some unimodular metric $\g_{\a\b}$. 
 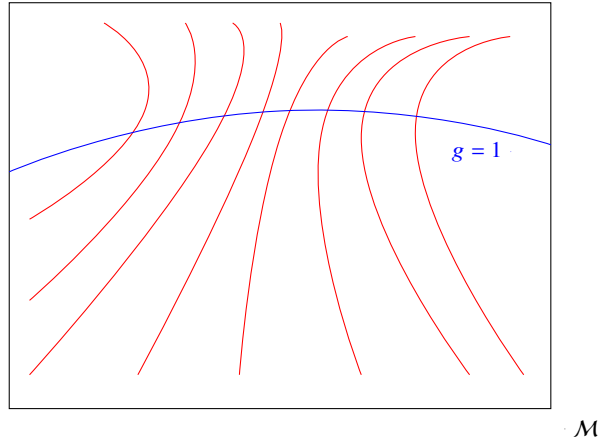
\begin{figure}[h!]\centering
  \begin{tikzpicture}[scale=1.8]
   \draw[black] (-1,-1) rectangle (3,2);
   \draw [red] plot [smooth, tension=1] coordinates {(-0.85,0.4) (0,1.2) (-0.3,1.85)};
   \draw [red] plot [smooth, tension=1] coordinates {(-0.85,-0.2) (0.2,1) (0.3,1.85)};
   \draw [red] plot [smooth, tension=1] coordinates {(-0.85,-0.75) (0.5,1) (0.65,1.85)};
   \draw [red] plot [smooth, tension=1] coordinates {(-0.05,-0.75) (0.8,1) (1,1.85)};
   \draw [red] plot [smooth, tension=1] coordinates {(0.7,-0.75) (1,1) (1.5,1.75)};
   \draw [red] plot [smooth, tension=1] coordinates {(1.6,-0.75) (1.3,1) (2,1.75)};
   \draw [red] plot [smooth, tension=1] coordinates {(2.4,-0.75) (1.6,1) (2.4,1.75)};
   \draw [red] plot [smooth, tension=1] coordinates {(2.8,-0.75) (2,1) (2.7,1.75)};
   \draw [blue] plot [smooth, tension=1] coordinates {(-1,0.75) (1,1.2) (3,0.95)};
   \filldraw [black] (3.1,-1.15) circle (0pt) node[anchor=west]{$\mathcal{M}$};
   \filldraw [blue] (2.7,0.9) circle (0pt) node[anchor=east]{$g=1$};
  \end{tikzpicture}   
   \caption{Weyl gauge orbits and gauge fixing for metrics $g=1$. } \label{fig:wg} 
  \end{figure}
  Then {\em any} Weyl rescaling of $\g_{\a\b}$ is a solution to \cref{emugbis}.
 The EM imply that in vacuum UG, spacetimes in this Weyl gauge are constant curvature spaces. Therefore, a detailed analysis of the collapse should be made to ascertain under what initial physical conditions a nonvanishing curvature is generated.
 Besides, many physical properties are {\em not} Weyl invariant. For example, flat space is conformally related to both de Sitter and anti de Sitter spacetimes, both being Petrov type 0.
 \par
 All solutions to the complete UG EM can be characterized by one unimodular solution $\g_{\a\b}$ in the Weyl gauge $g=1$ together with all their Weyl rescalings, $g_{\a\b}=g^{1\over 4} \g_{\a\b}$. The unimodular metric $\g$ is equivalent to the selection of a particular set of eigenvalues for one of the Petrov types (this being a Weyl invariant classification); see \cref{fig:wg} below.

 \section{Unimodular Cosmology.}\label{sec:Cosmology} 
 Consider the simplest cosmological model keeping the topology of the constant time surfaces flat. 
 The corresponding metric in the unimodular gauge, $g=1$, reads,
 \be
 ds^2=b(t)^{- 3/2}\,dt^2-b(t)^{1/2}\, \d_{ij}dx^i dx^j, \label{metcosmo} 
 \ee
 where $b=b(t)$, only depends on time. 
 The cosmic normalized four-velocity vector field, $u^\m u_\m=1$, is given explicitly by
 \be u^{\m}=\left(b^{3/4},0,0,0\right).\ee
 and the projector onto the 3-space is 
 \be
 \tensor{\Pi}{^\mu_\nu} \equiv \d^\m_\n-u^\m u_\n=\begin{pmatrix}0&0&0&0\\0&1&0&0\\0&0&1&0\\0&0&0&1\end{pmatrix}.
 \ee
 It is easy to check that this congruence is geodesic
 \be
 \label{c}u^\n\nabla_\n u^\m=0,
 \ee
 and the volume expansion reads
 \be \label{v1}\theta\equiv \nabla_\m u^\m=\frac{3}{4}b^{-1/4}\frac{db}{dt}.\ee

 The unimodular gauge has been used extensively, especially at the dawn of GR, in particular by Einstein himself and also by Schwarzschild \cite{Poisson}. 
 \par
 
 \par
 Once again, for this gauge, the EM in UG correspond to,
 \be 
 R_{\m\n}-\frac{1}{4}R\,g_{\m\n}=2\kappa^2 \left(T_{\m\n}-\frac{1}{4}\,T\,g_{\m\n}\right),\label{UGEM2} 
 \ee
 where $\kappa^2\equiv 8 \pi G$. 
 The scalar curvature corresponding to \cref{metcosmo} is
 \be\label{RR} R=-\frac{3}{8\sqrt{b}}\left[\left(\frac{db}{dt}\right)^2+4b\frac{d^2 b}{dt^2}\right].\ee
 Considering matter as a perfect fluid with a stress-energy tensor,
 \be
 T_{\m\n}\equiv \left(\r+p\right)u_\m u_\n-p g_{\m\n},
 \ee
 for which energy-momentum conservation, $\nabla_\n T^{\m\n}=0$ is then equivalent to 
 \be
 u^\m\nabla_\m{\r}+\left(\r+p\right)\theta=0.
 \ee
 Using the \cref{UGEM2} along with \cref{c}, Raychaudhuri's equation \cite{Raychaudhuri:1955} reduces to
 \be u^\m\nabla_\m{\theta}+\frac{1}{n-1}\theta^2+\s_{\a\b}\s^{\a\b}-\omega_{\a\b}\omega^{\a\b}+\frac{1}{n}R+\frac{2(n-1)}{n}\kappa^2(\r+p)=0.\ee
Using the explicit form for the Ricci scalar in \eqref{RR}, 
 \be R=-2u^\m\nabla_\m{\theta}-\frac{4}{3}\theta^2.\ee
 Assuming, as it is usual for simplicity, vanishing shear and rotation, $\s_{\a\b}=\omega_{\a\b}=0$, in the physical dimension $n=4$ 
 \be\label{Re} u^\m\nabla_\m{\theta}+3\kappa^2(\r+p)=0.\ee
 It is worth remarking that, due to the tracelessness of the EM, it is not possible to express $R$ in terms of $T$.
 \par
Using Ellis'clever trick \cite{Ellis}, one can define a length scale through
 \be \label{v2}\theta=\frac{3}{l}u^\m\nabla_\m l,\ee
 actually
 \be
 b\sim l^4.
 \ee
 Finally, one can write Raychaudhuri's equation \eqref{Re} like
 \be\label{le} u^\m u^\n\Big[ l\nabla_\n\nabla_\m{l}-\nabla_\m l\nabla_\n l\Big]+\kappa^2(\r+p)l^2=0.\ee
 In vacuum, $p=\r=0$ and Raychaudhuri's equation reduces to
 \be u^\m\nabla_\m{\theta}=0.\ee
 This is easy to calculate,
 \be u^\m\nabla_\m{\theta}=u^\n\nabla_\n\nabla_\m u^\m=-\frac{3}{16\sqrt{b}}\left[\left(\frac{db}{dt}\right)^2-4b\frac{d^2 b}{dt^2}\right]=0.\ee
  The vacuum EM for UG read,
 \be\left(\frac{db}{dt}\right)^2-4b\frac{d^2 b}{dt^2}=0.\label{152} \ee
The general solution to \cref{152} is given by either
 \be
 b=b_0,
 \ee
 (a constant) corresponding to flat space; or else by
 \be 
 b(t)=H_0^{\frac{4}{3}}\left(3t-t_0\right)^{\frac{4}{3}},
 \ee
 which corresponds to {\em de Sitter} \footnote{
 In unimodular coordinates, the maximally symmetric, constant curvature {\em de Sitter} spacetime reads
 \be\label{CC}
 ds^2=\left(\frac{dt}{3 H t}\right)^2-(3 H t)^{2/3} \d_{ij}\,dx^i dx^j
 \ee
 that is, precisely
 \be
 b(t)\sim \,t^{\frac{4}{3}}
 \ee} 
 space. Here $H_0\equiv 3\theta$ is the (constant) expansion rate, which for this solution is arbitrary, as there is no physical scale in the problem that can determine it. It is to be emphasized that this solution depends on two parameters, whereas the flat space solution depends only on one, being thus less generic.
 \par
 This could have been anticipated because the vacuum EM in UG are just Einstein spaces 
 \be
 R_{\m\n}=\frac{1}{4} R g_{\m\n}.
 \ee
 Flat space is just a pretty particular solution; non-zero constant curvature spacetimes \cite{Wolf} constitute the more generic ones.
 \\
 It is sometimes asserted that UG is equivalent to general relativity in the gauge $g=1$. However, this is not true, as shown in this paragraph. 
 
 First, the synchronous gauge cannot be reached with the residual gauge symmetry once in the unimodular gauge. 
 The best one can do, \cite{AlvarezFaedo}, is
 \be
 ds^2= a(t)dt^2-R^2(t) \d_{ij} dx^i dx^j,
 \ee
 where flat spatial sections have been chosen for simplicity. Einstein's equations with vacuum energy
 \be
 G_{\mu \nu}\equiv\tensor{R}{_\mu_\nu}-\frac{1}{2}R \tensor{g}{_\mu_\nu},\qquad T_{\m\n}\equiv \rho g_{\m\n},
 \ee
 read
 \bea
 G_{00}&&= 3 \frac{\dot{R}^2}{R^2}=\kappa^2 a \r,\nonumber\\
 G_{ij}&&=\frac{\dot{a}\dot{R}R-2 a \ddot{R} R-a \dot{R}^2}{a^2}\d_{ij}=-\kappa^2 \rho R^2 \d_{ij}.
 \eea
The unimodular gauge implies
 \be
 a R^6=1,
 \ee
 for which the straightforward solution is
 \bea
 &R=\left[R_0^3+\kappa\sqrt{\frac{\r}{3}}\,\left(t-t_0\right)\right]^{\frac{1}{3}},\nonumber\\
 &a=\left[R_0^3+\kappa\sqrt{\frac{\r}{3}}\,\left(t-t_0\right)\right]^{-2}.
 \eea
 
 This is just flat space in vacuum ($\r=0$) and an exponential expansion in the synchronous time
 \be
 \t=\int^t \frac{dx}{R_0^3+\kappa\sqrt{\frac{\r}{3}}\,\left(x-t_0\right)}=\frac{1}{ \kappa}\sqrt{\frac{3}{\r}}\log\,\left(R_0^3+\kappa\sqrt{\frac{\r}{3}}\,\left(t-t_0\right)\right),
 \ee
 in such a way that
 \be
 R(\t)=e^{\frac{\kappa}{3}\sqrt{\frac{\r}{3}}\left(\t-\t_0\right)},
 \ee
 and the exponential expansion disappears once $\r=0$.
 \\
 On the other hand, the unimodular gauge of GR is, of course, fully equivalent to the usual formulation of GR in comoving coordinates \cite{Weinberg, Bondi} where the metric reads
 \be\label{mc}
 ds^2= d\t^2-a(\t)^2 \sum\d_{ij} dx^i dx^j,
 \ee
 with a four-velocity
 \be u^\m=(1,0,0,0),\ee
 and 
 \be u^\n\nabla_\n{u^\m}=0.\ee
 In this case 
 \be 
 \theta=3\frac{1}{a}\frac{da}{dt}.
 \ee

Again, the only difference between GR and UG stems from the EM. 
In detail, now the EM are the usual Einstein ones
 \be 
 R_{\m\n}-\frac{1}{2}R\,g_{\m\n}=2\kappa^2 T_{\m\n},\label{EM} 
 \ee
 in this case, the scalar of curvature reads
 \be R=-\frac{6}{a^2}\left[\left(\frac{da}{dt}\right)^2+a\frac{d^2a}{dt^2}\right].\ee
Raychaudhuri's equation in comoving coordinates yields
 \be -3\left(\frac{da}{dt}\right)^2+2a^2\kappa^2\r=0.\ee
 In this case, the vacuum solution reduces to
 \be \frac{da}{dt}=0,\ee
 i.e., $\theta=0$, which is just flat spacetime.
 This is a subset of the unimodular result $\dot{\theta}=0$. 
  \section{Tree diagrams.}\label{sec:trees} 
 
  A natural question to ask at this stage is whether the S-matrix would be the same for UG as for GR. 
  The propagators and the vertices are pretty different in both theories, so the answer to this question is not immediate.
  In \cite{Alvarez:2016uog}, the calculation of the nonvanishing tree-level three, four, and five graviton amplitudes were compared with the diagrams in GR \cite{Benincasa,Berends,Cachazo}. 
  Complete agreement at the diagram level was found. 
  In this section, the key ingredients for that computation are presented.
   
   To obtain the Feynman rules for UG, consider the action \cref{UGact1}. 
   The propagator is obtained by inverting the second-order expansion\footnote{Around a flat background, $\partial^2\equiv\partial_\mu\partial_\nu \tensor{\eta}{^\mu^\nu}.  $}  of the Lagrangian - once properly gauge-fixed- presented in \cite{Alvarez2015}. This reads
   \begin{align}
   \nonumber {\cal L}&=\frac{1}{4}h^{\m\n}\partial^2 h_{\m\n}-\frac{1}{4 n}h \partial^2 h +\left(-f\partial^2 f +\frac{\alpha}{2}f\partial^2 h+\frac{\alpha}{2}h\partial^2 f\right)-\nonumber\\
   &-\frac{1}{2}\left(\partial_{\m}c'^{\;(0,0)}\partial^{\m}c'^{\;(0,0)}+2\left(\partial_{\n}h^{\n}_{\m}-\frac{1}{n}\partial_{\m}h\right)\partial^{\m}c'^{\;(0,0)}\right)
   \end{align}
   
   Writing the action as
   \begin{align}\label{form_operator}
   S=\int d^{n}x\; \Psi^{A}F_{AB}\Psi^{B},
   \end{align}
   where,
   \begin{align}
   F_{AB}=G_{AB}\partial^2+ J^{\m\n}_{AB}\partial_{\m}\partial_{\n},
   \end{align}
   and
   \begin{align}
   \Psi^{A}=\begin{pmatrix}
   h^{\m\n}\\
   f\\
   c' \\
   \end{pmatrix}.
   \end{align}  
   Defining the auxiliary tensors
   \begin{align}
   {\cal P}_{\m\n\r\s}^{\a\b}&=\frac{1}{4}\left(\eta_{\m\r}\delta_{\n}^{(\a}\delta_{\s}^{\b)}+\eta_{\m\s}\delta^{(\a}_{\n}\delta^{\b)}_{\r}+\eta_{\n\r}\delta^{(\a}_{\m}\delta^{\b)}_{\s}+\eta_{\n\s}\delta^{(\a}_{\m}\delta^{\b)}_{\r}\right),\\
   {\cal K}_{\m\n\r\s}^{\a\b}&=\frac{1}{2}\left(\eta_{\m\n}\delta^{(\a}_{\r}\delta^{\b)}_{\s}+\eta_{\r\s}\delta^{(\a}_{\m}\delta^{\b)}_{\n}\right).
   \end{align}
 The different matrices involved read
   \begin{align}
   & G_{AB}=\begin{pmatrix}
   -\frac{1}{4}\left(\frac{1}{4}{\cal K}_{\m\n\r\s}^{\a\b}-{\cal P}_{\m\n\r\s}^{\a\b}\right)\eta_{\a\b} \quad&\frac{\a}{2}\eta_{\m\n}& -\frac{1}{8}\eta_{\m\n}\\
   \frac{\a}{2} \eta_{\r\s} & -1 &0\\
   -\frac{1}{8}\eta_{\r\s}&0&\frac{1}{2}
   \end{pmatrix},\label{matrix}\\
   & J^{\a\b}_{AB}=\begin{pmatrix}
   0&0&\frac{1}{4}\left(\delta^{\a}_{\m}\delta^{\b}_{\n}+\delta^{\a}_{\n}\delta^{\b}_{\m}\right)\\
   0&0&0\\
   \frac{1}{4}\left(\delta^{\a}_{\r}\delta^{\b}_{\s}+\delta^{\a}_{\s}\delta^{\b}_{\r}\right)&0&0
   \end{pmatrix}.
  \end{align}
   For the gauge choice of \cite{Alvarez2015}, the graviton propagator in UG reads
   \bea
   & P^\text{UG}_{\m\n,\r\s}=\dfrac{1}{2k^2}\left(\eta_{\m\s}\eta_{\n\r}+\eta_{\m\r}\eta_{\n\s}\right)-\dfrac{1}{k^2}\dfrac{\a^2n^2-n+2}{\a^2 n^2(n-2)}\eta_{\m\n}\eta_{\r\s}+\nonumber\\
   &+\dfrac{2}{n-2}\left(\dfrac{k_\rho k_\s \eta_{\m\n}}{k^4}
   +\dfrac{k_\m k_\n \eta_{\r\s}}{k^4}\right)-\dfrac{2n}{n-2}\dfrac{k_{\m}k_{\n}k_{\r}k_{\s}}{k^6}\label{propagatorug}.
   \eea 
   
   Recall that the usual GR graviton propagator in the de Donder gauge,
   \be P_{\m\n\r\s}^{\text{GR}}=\dfrac{i}{2k^2}\left(\eta_{\m\s}\eta_{\n\r}+\eta_{\m\r}\eta_{\n\s}-\dfrac{2}{n-2}\eta_{\m\n}\eta_{\r\s}\right),
   \ee
   has only simple poles at $k^2=0$. 
   In the UG propagator,
   there appear additional double and triple poles. 
   This is a technical complication that impedes \textit{a priori}, the application of the techniques \cite{Britto} to reduce the computation of the diagrams. Furthermore, it is possible to show that ,  if the Newtonian potential is to be obtained in the non-relativistic limit, no gauge choice in UG can yield a propagator of the form
   \begin{align} 
   P_{\m\n\r\s}&=\dfrac{i}{2k^2}\left(\eta_{\m\s}\eta_{\n\r}+\eta_{\m\r}\eta_{\n\s}-f_1(k^2)\,\eta_{\m\n}\eta_{\r\s}\right)+f_2(k^2)
   (k_\rho k_\rho \eta_{\m\n}+k_\m k_\n \eta_{\r\rho})+\nonumber\\
   &\qquad+f_3(k^2)\,k_{\m}k_{\n}k_{\r}k_{\rho},
   \end{align}
   with $f_3(k^2)$ having no pole at $k^2=0$. 
  The triple pole term in (\ref{propagatorug}) is needed to recover said potential.
   
\begin{figure}[h!]
   \centering 
   \begin{tikzpicture}[scale=1.5]
    \filldraw [black] (-0.5,0.75) circle (0pt) node[anchor=south]{\large$p_1$};
    \filldraw [black] (-0.5,-0.8) circle (0pt) node[anchor=north]{\large$p_2$};
    \filldraw [black] (0.8,0.12) circle (0pt) node[anchor=south]{\large$q$};
    \filldraw [black] (2.1,0.7) circle (0pt) node[anchor=south]{\large$p_4$};
    \filldraw [black] (2.1,-0.8) circle (0pt) node[anchor=north]{\large$p_3$};
    \draw[spring] (-1,1) -- (0,0);
    \draw[spring] (-1,-1) -- (0,0);
    \draw[spring] (0,0) -- (1.7,0);
    \draw[spring] (1.7,0) -- (2.7,1);
    \draw[spring] (1.7,0) -- (2.7,-1);
   \end{tikzpicture}
    \caption{$s$ channel.}\label{fig:3} 
    \end{figure}
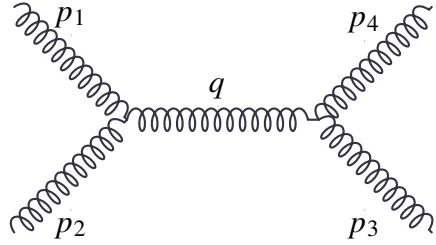

  The three-and-four graviton vertices are needed to calculate the four and five-point amplitudes. These are obtained from the second and third-order expansion of the Lagrangian. The results can be presented in a compact form via the parameter $n$. For GR vertices $n=2$ while for UG $n=4$. 
  Using the {\em all incoming momenta} convention in the diagrams in \cref{fig:1,fig:2,fig:3}, the three-graviton vertex reads
   \begin{align}
   V&^{\m\n,\r\rho,\a\b}_{(p1,p2,p3)}=i 2\kappa ~{\cal S}\Bigg\{ \frac{(2+n)(p_1.p_2) }{2n}\left[ \frac{\eta^{\a\b} \eta^{\m\n} \eta^{\r\rho}}{ n^2}- \frac{ 2 \eta^{\a\r} \eta^{\b\rho} \eta^{\m\n}}{n} - \frac{ \eta^{\a\b} \eta^{\m\r} \eta^{\n\rho}}{2 +n} \right]+\nonumber\\
   & + \frac{2 \eta^{\b\n} \eta^{\r\rho} p_1^{m} p_2^{\a}}{n} + \tfrac{1}{2} \eta^{mr} \eta^{\n\rho} p_1^{\a} p_2^{\b} - \frac{\Bigl(2 + n\Bigr) \eta^{\m\n} \eta^{\r\rho} p_1^{\a} p_2^{\b}}{2 n^2} - 2 \eta^{\b\rho} \eta^{\n\r} p_1^{\a} p_2^{\m} - \eta^{\a\n} \eta^{\b\rho} p_1^{\r} p_2^{\m} +\nonumber\\
   & + \frac{\eta^{\a\b} \eta^{\n\rho} p_1^{\r} p_2^{\m}}{n} + \frac{2 \eta^{\b\m} \eta^{\r\rho} p_1^{\a} p_2^{\n}}{n} - \frac{2 \eta^{\a\b} \eta^{\r\rho} p_1^{\m} p_2^{\n}}{n^2} + \frac{2 \eta^{\a\m} \eta^{\b\n} p_1^{\rho} p_2^{\r}}{\n}+ (p_1.p_2) \eta^{\a\n} \eta^{\b\rho} \eta^{\m\r}\Bigg\}.
   \end{align}
   The four-graviton vertex, corresponds to,
   \begin{align}
   V&^{\m\n,\r\s,\a\b,\eta\l}_{(p1,p2,p3,p4)}=i 2s\kappa~ {\cal S}\Bigg\{ \frac{(2+n )(p_3.p_4)}{n^2}\left[\frac{ g^{\m\n} g^{\r\s} g^{\a\b} g^{\eta\lambda}}{4 n^2} - \frac{ g^{\m\r} g^{\a\b} g^{\eta\lambda} g^{\n\s}}{4 n} + g^{\m\r} g^{\a\b} g^{\eta\s} g^{\n\lambda}\right] -\nonumber\\
   &- \tfrac{1}{2} (p_3.p_4) g^{\m\eta} g^{\r\lambda} g^{\a\n} g^{\s\b} + \frac{\Bigl(2 + n\Bigr) (p_3.p_4) g^{\m\eta} g^{\r\a} g^{\n\lambda} g^{\s\b}}{2 n^2} - \frac{\Bigl(2 + n\Bigr) (p_3.p_4) g^{\m\n} g^{\r\eta} g^{\a\b} g^{\s\lambda}}{ n^3} -\nonumber\\
   &- \frac{(p_3.p_4) }{n} \left[\frac{ g^{\m\n} g^{\r\s} g^{\a\eta} g^{\b\lambda}}{4 n} + g^{\m\n} g^{\r\eta} g^{\a\s} g^{\b\lambda} - n g^{\m\r} g^{\a\n} g^{\eta\s} g^{\b\lambda} + \frac{g^{\m\r} g^{\a\eta} g^{\n\s} g^{\b\lambda}}{4 }\right]+\nonumber\\
   &+ g^{\m\eta} g^{\a\s} g^{\b\lambda} p_3^{\n} p_4^{\r} + \frac{2+n}{2n^2}\left[ g^{\m\r} g^{\a\b} g^{\eta\lambda} p_3^{\s} p_4^{\n} - \frac{ g^{\m\n} g^{\a\b} g^{\eta\lambda} p_3^{\r} p_4^{\s}}{n}+ 2 g^{\m\a} g^{\eta\lambda} g^{\n\b} p_3^{\r} p_4^{\s}\right]-\nonumber\\
   &- \tfrac{1}{2} g^{\m\r} g^{\a\eta} g^{\b\lambda} p_3^{\s} p_4^{\n} + \frac{g^{\m\n} g^{\a\eta} g^{\b\lambda} p_3^{\r} p_4^{\s}}{2 n} - g^{\m\a} g^{\eta\n} g^{\b\lambda} p_3^{\r} p_4^{\s} - 2\frac{g^{\m\a} g^{\r\b} g^{\eta\lambda} p_3^{\n} p_4^{\s}}{n} +\nonumber\\
   &+ 2g^{\m\r} g^{\a\lambda} g^{\eta\s} p_3^{\n} p_4^{\b} - 2\frac{g^{\m\r} g^{\a\s} g^{\eta\lambda} p_3^{\n} p_4^{\b}}{n} + 2g^{\m\eta} g^{\r\lambda} g^{\a\n} p_3^{\s} p_4^{\b} - 2\frac{g^{\m\n} g^{\r\eta} g^{\a\lambda} p_3^{\s} p_4^{\b}}{n} +\nonumber\\
   &+ 2\frac{g^{\m\n} g^{\r\a} g^{\eta\lambda} p_3^{\s} p_4^{\b}}{n^2} - 2\frac{g^{\m\eta} g^{\r\a} g^{\n\lambda} p_3^{\s} p_4^{\b}}{n} + \frac{g^{\m\n} g^{\r\s} g^{\a\eta} p_3^{\lambda} p_4^{\b}}{2 n^2} - \frac{g^{\m\n} g^{\r\eta} g^{\a\s} p_3^{\lambda} p_4^{\b}}{ n} \nonumber\\
   &+ g^{\m\r} g^{\a\n} g^{\eta\s} p_3^{\lambda} p_4^{\b} - \frac{g^{\m\r} g^{\a\eta} g^{\n\s} p_3^{\lambda} p_4^{\b}}{2 n} - 2\frac{g^{\m\a} g^{\eta\s} g^{\n\b} p_3^{\r} p_4^{\lambda}}{n} - \frac{g^{\m\n} g^{\r\s} g^{\a\b} p_3^{\eta} p_4^{\lambda}}{ n^3} +\nonumber\\
   &+ \frac{g^{\m\r} g^{\a\b} g^{\n\s} p_3^{\eta} p_4^{\lambda}}{ n^2} - 2\frac{g^{\m\r} g^{\a\s} g^{\n\b} p_3^{\eta} p_4^{\lambda}}{n} + 2\frac{g^{\m\n} g^{\r\a} g^{\s\b} p_3^{\eta} p_4^{\lambda}}{n^2} - \nonumber\\
   &- 2\frac{g^{\m\r} g^{\a\b} g^{\eta\s} p_3^{\n} p_4^{\lambda}}{n} + 2\frac{g^{\m\n} g^{\r\eta} g^{\a\b} p_3^{\s} p_4^{\lambda}}{n^2}\Bigg\}.
   \end{align}
   
   Where $\mathcal{S}$ is a shorthand
   for a double symmetrization, namely
   \begin{enumerate}
   \item A summation over all momentum-index combinations ($p_1,{\m\n}$; $~p_2,\r\s$; $~;p_3,\a\b$; $~p_4,\eta\l$).
   \item A symmetrization of each pair on indices $\m\n$, $\r\s$, $\a\b$, $\eta\l$. 
   \end{enumerate}

   The fact that UG perturbatively expanded around Minkowski space is Lorentz invariant, and the graviton polarizations are the same as in GR means, by the standard analysis -- see \cite{Elvang:2015rqa}, that the on-shell three-point amplitudes vanish on-shell for real momenta. 
  Since the little group scaling operates in UG precisely in the same manner as in GR, it is plain that for conserved complex momenta, the on-shell nonvanishing three-point amplitudes are the same in both theories but, perhaps, for a global constant. 
  By explicit computation of the corresponding diagrams, it can be shown that the constant in question is the same in both theories, as it becomes the fact that the classical Newton constant is also the same in both theories see \cref{sec:Sources}.
   \\
  Having clarified the three-graviton diagrams, one can further consider four-graviton tree amplitudes.
  Since the pure four-vertex diagram vanishes, only three types of diagrams involve four external gravitons. These are the well-known $s$, $t$, and $u$ channels, see \cref{fig:1,fig:2,fig:3}. 

   \begin{figure}[h!]
    \centering
    \begin{minipage}{0.47\linewidth}
    \centering
    \begin{tikzpicture}[scale=2.3]
      \filldraw [black] (-0.65,0.5) circle (0pt) node[anchor=south]{\large$p_1$};
      \filldraw [black] (-0.65,-0.6) circle (0pt) node[anchor=south]{\large$p_2$};
      \filldraw [black] (0.12,0) circle (0pt) node[anchor=west]{\large$q$};
      \filldraw [black] (0.65,0.5) circle (0pt) node[anchor=south]{\large$p_4$};
      \filldraw [black] (0.65,-0.6) circle (0pt) node[anchor=south]{\large$p_3$};
      \draw[spring] (0,0.5) -- (0,-0.5);
      \draw[spring] (-1,1) -- (0,0.5);
      \draw[spring] (0,0.5) -- (1,1);
      \draw[spring] (-1,-1) -- (0,-0.5);
      \draw[spring] (0,-0.5) -- (1,-1);
     \end{tikzpicture}
    \caption{$u$ channel.} \label{fig:1} 
    \end{minipage}
    \begin{minipage}{0.47\linewidth}
   \centering
    \centering
    \begin{tikzpicture}[scale=2.3]
      \filldraw [black] (-0.65,0.5) circle (0pt) node[anchor=south]{\large $p_1$};
      \filldraw [black] (-0.65,-0.6) circle (0pt) node[anchor=south]{\large$p_2$};
      \filldraw [black] (0.12,0) circle (0pt) node[anchor=west]{\large$q$};
      \filldraw [black] (0.65,0.5) circle (0pt) node[anchor=south]{\large$p_3$};
      \filldraw [black] (0.65,-0.6) circle (0pt) node[anchor=south]{\large$p_4$};
      \draw[spring] (0,0.5) -- (0,-0.5);
      \draw[spring] (-1,1) -- (0,0.5);
      \draw[spring] (0,0.5) -- (1,1);
      \draw[spring] (-1,-1) -- (0,-0.5);
      \draw[spring] (0,-0.5) -- (1,-1);
     \end{tikzpicture}
    \caption{$t$ channel.}\label{fig:2} 
    \end{minipage}
   \end{figure}

   Following, \cite{Alvarez2016-2} the polarization tensors for the gravitons are written in terms of those of the gluon: 
  
   \be \epsilon^-_{\m\n}=\epsilon^-_\m\epsilon^-_\n\longrightarrow\epsilon^-_{a\dot{a},b\dot{b}}=\epsilon^-_{a\dot{a}}\epsilon^-_{b\dot{b}}~~\text{and}~~\epsilon^+_{a\dot{a},b\dot{b}}=\epsilon^+_{a\dot{a}}\epsilon^+_{b\dot{b}}.\ee
   The resulting amplitudes are then:
   \begin{align}
   {\cal A}_s(1^-2^-;3^+4^+)&=\e_1^{-\m_1}\e_1^{-\n_1}\e_2^{-\m_2}\e_2^{-\n_2}V^{\m_1\n_1,\m_2\n_2,\a\b}_{(p1,p2,q)}P_{\a,\b,\r,\s}V^{\r\s,\m_3\n_3,\m_4\n_4}_{(p,p3,p4)}\e_3^{-\m_3}\e_3^{-\n_3}\e_4^{-\m_4}\e_4^{-\n_4}\nonumber\\
   &=- \frac{i 4 \kappa^2 (e_1.p_2)^2 (e_2.e_3)^2 (e_4.p_2)^2}{s^2},\label{udiagr} \\
   {\cal A}_t(1^-3^+;2^-4^+)&=\e_1^{-\m_1}\e_1^{-\n_1}\e_3^{-\m_3}\e_3^{-\n_3}V^{\m_1\n_1,\m_3\n_3,\a\b}_{(p1,p3,q)}P_{\a,\b,\r,\s}V^{\r\s,\m_2\n_2,\m_4\n_4}_{(p,p2,p4)}\e_2^{-\m_2}\e_2^{-\n_2}\e_4^{-\m_4}\e_4^{-\n_4}\nonumber\\
   &=0,\label{tdiagr} \\
   {\cal A}_u(1^-4^+;2^-3^+)&=\e_1^{-\m_1}\e_1^{-\n_1}\e_4^{-\m_4}\e_4^{-\n_4}V^{\m_1\n_1,\m_4\n_4,\a\b}_{(p1,p4,q)}P_{\a,\b,\r,\s}V^{\r\s,\m_2\n_2,\m_3\n_3}_{(p,p2,p3)}\e_2^{-\m_2}\e_2^{-\n_2}\e_3^{-\m_3}\e_3^{-\n_3}\nonumber\\
   &=- \frac{i 4 \kappa^2 (e_1.p_2)^2 (e_2.e_3)^2 (e_4.p_2)^2}{u^2}. \label{sdiagr} 
   \end{align}
   In \cref{udiagr,tdiagr,sdiagr},  as usual, $s=p_1+p_2$ and $u=p1+p3$.
   These amplitudes are \textit{diagram to diagram} the same as those for GR.
   \\
   This equivalence is maintained for the tree diagrams with five external gravitons, as shown in \cite{Alvarez:2016uog}.
   \\
   To summarize, it has been shown that the Maximal Helicity Violating (MHV), three, four, and five graviton tree amplitudes give the same contribution in GR and UG. Moreover, this result holds for each diagram independently and not only for the total amplitude.
   Therefore, at least at tree level, and with three, four, or five external legs, the MHV contribution to the S matrix for pure gravity without coupling to other fields is the same in both theories.
   
   A remarkable fact is that all the terms that involve the double and triple poles in the propagator of UG \eqref{propagatorug} do not contribute to any diagram.
   \par
   This cancelation is not trivial, and it came as a surprise in the original work. Indeed, in UG, one obtains the following non-zero result
   \be
   k_{\a} k_{\b} V^{\m\n,\r\s,\a\b}_{(p,q,k)}\, \e_{1\,\m\n}(p)\,\e_{2\,\r\s}(q) = i\kappa\,(p\cdot q)\,(p\cdot \e_{2})(q\cdot\e_{1})\, (\e_{1}\cdot\e_{2}),
   \ee
   when $k=-p-q$ is off-shell and the polarizations with well-defined helicity $\e_{1\,\m\n}(p)=\e_{1\,\m}(p)\e_{1\,\n}(p)$ and $\e_{2\,\r\s}(q)=\e_{2\,\r}(q)\e_{2\,\s}(q)$ are arbitrary. 
   On the other hand, the computation of the corresponding object in GR yields a vanishing result as a consequence of \textit{Diff} invariance. It could be that the gauge invariance of UG, and the fact that the only physical degrees of freedom in perturbative UG are gravitons, are the ones to be held responsible, at least partially, for this cancellation.

   \section{One-Loop Unimodular Gravity.}\label{sec:quantization}
   Until this point, the comparison of UG and GR has considered the EM and their solutions and tree diagrams. 
   Both UG and GR propagate the same number of degrees of freedom.
   Furthermore, the tree diagrams explored in \cref{sec:trees} coincide with those of GR despite the notable differences for the propagators in both theories.
   The reader may then be inclined to think that it must be the case that any tree-level computation might be equivalent for both theories. Computations in \cref{sec:trees} explicitly show that this is far from trivial. 
   
  In any case, any potential equivalence at the tree level need not extrapolate to loop-level computations. 
   In the path integral approach to the quantization of UG, the path integral measure must incorporate the fact that the gauge group is not \textit{Diff} but \textit{WTDiff}, as discussed for example in \cite{Ardon2017}. 
   
   Therefore, while on-shell states match for both UG and GR, since the two theories do not share gauge groups, while GR gravitons running on loops being off-shell need not be traceless, UG gravitons are traceless even inside loops.
   Simply put, loops run over different states in each case. 
   Therefore, in perturbation theory, at least \textit{a priori}, one lacks a reason to expect the cancellations of these differences at higher orders. 
   From the reasons above, one could expect potential differences to arise between UG and GR at the one-loop level. 
   
   In the last part of this chapter, a particular formalism\footnote{This choice is by no means unique, and there is an ongoing discussion on whether different approaches to the quantization might yield different results \cite{UGreport}. 
   Despite its interest, this possible inequivalence will not be further discussed here. } that allows dealing with the involved gauge sector of UG in one-loop computations is presented. 
   
   Consider the {background field expansion}. 
   Once again, the metric is split field into the sum of a \textit{classical} and a \textit{quantum} part,
   \begin{equation}
   \tensor{g}{_\mu_\nu}=\tensor{\overline{g}}{_\mu_\nu} +\kappa\tensor{h}{_\mu_\nu}. 
   \end{equation}
   Additionally, assume the measure of the path integral is shift-invariant, i.e., 
   \begin{equation}
   \int_{}\left[\mathcal{D} \tensor{g}{_\mu_\nu} \right]\, \,\operatorname{exp}\left[i \,S[g]+T\cdot g \right]=\int_{}\left[\mathcal{D} \tensor{h}{_\mu_\nu} \right]\, \,\operatorname{exp}\left[i \,S[h]+T\cdot g\right],
   \end{equation}
   where $T\cdot g$ represents the sources that are added to the partition function to obtain field expectation values by functional differentiation,
   \begin{equation}
   T\cdot g=\int_{}^{}d^{4}x \sqrt{g}\,\tensor{T}{^\mu^\nu} \tensor{g}{_\mu_\nu}.
   \end{equation}
   \indent The beauty of this method is that it allows fixing the gauge of the quantum part while preserving the gauge invariance of the background. 
   Then all computations are invariant under gauge transformations of the background\footnote{Is in this sense that this approach is \textit{Covariant}.}. The same happens for the counterterms.
   
   As it happens for other gauge theories, gauge freedom introduces an indeterminacy in the path integral. 
   To resolve this, one often uses the DeWitt-Feynman-Fadeev-Poppov method \cite{DeWitt1967,Faddeev1967}. 
   This adds to the original action, a part coming from the gauge fixing and certain ghost content representing the jacobian of the gauge fixing condition. 
   \begin{equation}
   S\rightarrow S+S_{\mbox{\tiny G.F.}}+S_{ \mbox{\tiny Gh}},\label{gfa} 
   \end{equation}
   finding a gauge fixing condition for the quantum \textit{WTDiff} that preserves the background's \textit{WTDiff} invariance is an open problem \cite{Herrero2020}. 
   Invariance under \textit{WTDiff} implies, in particular, that the generators of the \textit{TDiff} sector have to be transverse. 
   
   This is the main reason for the more involved nature of the gauge sector of UG. Nevertheless, these complications can be tackled at the expense of introducing new\footnote{Not present for GR.} bosonic ghost fields.
   
   To resolve the technicalities discussed above, one can consider the BRST approach as an alternative to the usual gauge-fixing procedure.
   The basic ideas for the unfamiliar reader are summarized below. 
   It was found that the gauge fixed action \cref{gfa} is invariant under a BRST transformation \cite{BRS,T}.
   This implies that the physical content of the gauge theory is given by the cohomology class of the BRST nilpotent operator $\mathfrak{s}$. 
   In the case at hand one can split,
   \begin{align}
   &\mathfrak{s}=\mathfrak{s}_W+\mathfrak{s}_D,\qquad \qquad \quad\mbox{where}\\
   &\mathfrak{s}\tensor{\overline{g}}{_\mu_\nu} =0,\\
   &\mathfrak{s}\tensor{h}{_\mu_\nu} =\mathfrak{s}_D\tensor{h}{_\mu_\nu} +\mathfrak{s}_W\tensor{h}{_\mu_\nu} =\mathcal{L}_{c^\mu}(\tensor{\overline{g}}{_\mu_\nu}+\tensor{h}{_\mu_\nu} )+2c(\tensor{\overline{g}}{_\mu_\nu}+\tensor{h}{_\mu_\nu} ).
   \end{align}
  The action is given in terms of the original lagrangian density by,
   \begin{equation}
   S_{\mbox{\tiny BRST}}=\int_{}^{}d^{4}x\, \mathcal{L}+\mathfrak{s}\Psi.\label{BRSTACT} 
   \end{equation}
   Now, because the generator of the volume-preserving diffeomorphism is transverse, so must be the associated ghost field $c^\mu$,
   \begin{equation}
   D_\mu c^\mu=0,
   \end{equation}
   where the transversality condition is given by the Weyl covariant derivative.
   
   The easiest way to implement this is using a projector \cite{Alvarez2015,Herrero2020} $\Pi^\mu_\nu$ acting on an unrestricted $c^\nu$. 
   This introduces a $U(1) $ invariance,
   \begin{equation}
   c^\nu \to D^\nu f.
   \end{equation} 
   A general treatment of such an algebra requires introducing the BV quantization techniques \cite{BV1981,Lyakhovich}. 
   Nevertheless, in this case, the procedure can be carried out, to the one-loop level, by enlarging the ghost content of the theory \cite{Ohta1,Ohta2,Ohta3}.
   Let us now give the guidelines to do so.  
   
   The first step is to include the necessary set of anti-ghost and auxiliary fields that close the algebra. 
   \begin{align}
   &\tensor{h}{_\mu_\nu^{(0,0)}},\;c_\mu^{(1,1)},\;b_\mu^{(1,-1)},\;f_\mu^{(0,0)},\;\phi^{(0,2)} \nonumber \\
   &\pi^{(1,-1)},\;\pi'^{(1,1)},\;\overline{c}^{(0,-2)},\; c'^{(0,0)}\nonumber\\
   &c^{(1,1)},\;b^{(1,-1)},\;f^{(0,0)}, \label{fields} 
   \end{align}
   where in the label $(n,m)$, $n$ denotes the Grasmann number (modulo 2) and $m$ is the ghost number.
   The first line corresponds to the physical graviton and the ghost fields that one would naively need for \textit{Diff}. In adition $\phi$ corresponds to the $U(1)$ transformation. 
   The second line has the ghost content to fix that $U(1)$, and the last line corresponds to the Weyl invariance.
   
   This is enough to make $\mathfrak{s}$ nil-potent to the one-loop level, ensuring BRST invariance of the total action.
   For completeness, the counterterm reads,
   \begin{align}
   &S_{\mbox{\tiny BRST}}^{\mbox{\tiny TDiff}}+S_{\mbox{\tiny BRST}}^{\mbox{ \tiny Weyl}}=\int_{}^{}d^{n}x\; b^\mu\Big(\Box^2c_\mu^{(1,1)}-2 \tensor{R}{_\mu_\rho} \nabla^\rho\nabla^\nu c_\nu^{(1,1)}-\Box \tensor{R}{_\mu^\rho} c_\rho^{(1,1)}-\nonumber\\
   &\;-2\nabla_\sigma \tensor{R}{_\mu^\rho} \nabla^\sigma c_\rho^{(1,1)}-\tensor{R}{_\mu_\rho} \tensor{R}{^\rho^\nu} c_\nu^{(1,1)}\Big)-\overline{c}^{(0,-2)}\Box\phi^{(0,2)}+\pi^{(1,-1)}\Box\pi'^{(1,1)}-\nonumber\\
   &\;-\frac{1}{\rho_1}\left(F_\mu F^\mu+\nabla_\mu c'^{(0,0)}\nabla^\mu c'^{(0,0)}+2F_\mu\nabla^\mu c'^{(0,0)}\right)-f^{(0,0)}\Box f^{(0,0)}+\frac{\alpha}{2} f^{(0,0)}\Box h+\nonumber\\
   &\;+\frac{\alpha}{2}h \Box f^{(0,0)}+2n \alpha b^{(1,-1)}\Box c^{(1,1)},\qquad \mbox{where}\quad F_\mu\equiv \nabla^\nu \tensor{h}{_\mu_\nu} -\frac{1}{n}\nabla_\mu h.
   \end{align}
   Once the ghost content issue is solved, one can apply the Schwinger-DeWitt proper time expansion\footnote{See \cite{Covariant} for an introduction to these techniques. } to calculate the divergences of UG \cite{Alvarez2015}. 
  Let us present here the main ideas. The sum of the divergent parts \footnote{The result in \cref{188}, which uses the gauge choice of \cite{Alvarez2015}, could seem to imply that one cannot make the terms proportional to $R$ and $R^2$ vanish, but as \cref{189} shows, one can remove these terms employing a two-parameter gauge fixing \cite{Kallosh1978}.  }  corresponds to \cite{Alvarez2015}: 
   \begin{align}
   W_{\infty}=\dfrac{1}{16\pi^2}\frac{1}{n-4} \int d^{n}x\Bigg(&\frac{119}{90}R_{\m\n\a\b}R^{\m\n\a\b}+\left( \frac{1}{6 \alpha^2}-\frac{359}{90} \right) R_{\m\n}R^{\m\n}+\nonumber\\
   &+\frac{1}{72} \left(22 - \frac{3}{\alpha^2}\right) R^{2}\Bigg).\label{188} 
   \end{align}
   
   Start by focusing on the issue of on-shell renormalizability. It is known that although GR is one-loop finite in the absence of a CC, this property is lost in its presence. The on-shell counterterm, in this case, was obtained in \cite{Christensen1980}, and it amounts to a renormalization of the CC and is proportional to
   \be
   W^{GR}_\infty\equiv {1\over 16\pi^2 (n-4)}\int \sqrt{g} d^4 x~\left({53\over 45}~R_{\m\n\a\b}R^{\m\n\a\b}-{1142\over 135}\Lambda^2\right).\label{189} 
   \ee
   
   Since the main attractive feature of UG is precisely the different r\^ole that the CC plays in contrast to GR, it is interesting to see what happens here with the renormalization group flow when the counterterm is taken to be on-shell so that all external legs correspond to physical states. 
   In that case, the EM for the $g=1$ fixed background are the traceless Einstein equations
   \begin{align}
   R_{\m\n}-\frac{1}{4}R g_{\m\n}=0,
   \end{align}
   which, altogether with Bianchi identities, imply the following for the operators appearing in the effective action
   \begin{align}
   &R_{\m\n\a\b}R^{\m\n\a\b}=E_{4},\\
   &R_{\m\n}R^{\m\n}=\frac{1}{4}R^{2},\\
   &R=\text{constant}.
   \end{align}
   The first line is nothing more than the Gauss-Bonet theorem when considering the EM. $E_4$ is thus the Pfaffian (the Euler density), whose integral gives the Euler characteristic of the manifold.
   
   By using these, the on-shell effective action can be cast in the form
   \begin{align}
   W_{\infty}^{\text{on-shell}}=\dfrac{1}{16\pi^2}\frac{1}{n-4} \int d^{n}x\left(\frac{119}{90} E_{4}-\frac{83}{120} R^2\right).
   \end{align}
   
   The contribution of the CC to the effective action is a non-dynamical quantity. Furthermore, this contribution does not couple to the metric because the $\sqrt{g}$ factor in the integration measure is absent. This implies one can disregard this term since it will not contribute to any correlator involving physical fields. 
   One can therefore conclude that in this case, there is no renormalization of the CC, and its peculiar status in UG is preserved through quantum corrections.
   
   Indeed, this effect is not specific to one-loop computations. The bare value of the CC is protected, and quantum corrections do not modify it. 
   \par
   It could be thought that this effect is just a gauge artifact of our background choice $\gamma=1$. However, it can be easily argued that this is not the case. As previously commented in this work, to obtain the effective action for an arbitrary background from the one with a unimodular background metric, it is enough to make a change of variables so that
   \begin{align}
   \gamma_{\m\n}=g^{-\frac{1}{n}}g_{\m\n}.
   \end{align}
   
   This transformation is available as long as a conformal anomaly is not generated. 
   This has been argued to be the case since a regularization scheme exists in which the anomaly vanishes \cite{Alvarezhv}. 
   Although this argument bears similarity to the classic work of \cite{Englert}, here it is applied to a tautological symmetry (what Duff \cite{Duff1994}  calls \textit{pseudo-Weyl} invariance) because there is a field redefinition in which the symmetry disappears. 
   In this case, the consensus is, \cite{Duff1994}, that there is no anomaly. 
Nevertheless,  the final word on this point is still to be said. 
  
   When doing this, one can see that the real reason for the CC not being renormalized is indeed the presence of Weyl invariance in the formalism, which protects the appearance of any mass scale in the effective action and, as a consequence, in the expectation value of the EM. 
   Therefore, the argument holds, and the CC is protected and fixed to its bare value all along the renormalization group flow and at any loop order.
   \section{Summary.}\label{sec:conclusions}
   This work is devoted to an introduction to Unimodular Gravity (UG).
   \par
    The primary motivation to consider UG is the r\^ole of the Cosmological Constant in this theory. The main difference in this respect is that a constant vacuum energy does not weigh at all and does not induce a Cosmological Constant. 
    
    The most direct path to UG is to start by demanding the unitarity of the linear,  spin-two field theory. There are only two solutions: one is the Fierz-Pauli theory, and the other is UG.
    
    After this, a general discussion on the fully non-linear EM and their solutions has been presented. 
    The coupling to matter sources is discussed in detail in \cref{sec:Sources}. The static potential agrees precisely with the one of General Relativity (GR). 
    It can also be argued at the non-linear level that there is an equivalence of sorts between UG and GR with some Cosmological Constant, which is determined by the boundary conditions of the equations of motion and is not related to the presence of any vacuum energy.
  
    It has even been claimed that the equivalence from the classical perspective can be generalized to higher-derivative theories of gravity, see \cite{UGreport}.
    However, Birkhoff's theorem is not valid in UG for a variety of reasons; \cref{sec:Birkhoff}. This leads, in particular, to exponentially expanding vacuum solutions in cosmology \cref{sec:Cosmology}.
    
    The last part of this chapter deals with a perturbative formulation of the path integral. Some improvement over GR in the ultraviolet is expected here because there is no integration over the conformal factor in UG. However, this fact is obscured in the Weyl invariant formulation presented in this work. Moreover, BRST symmetry is somewhat tricky due to the interference between Weyl and \textit{TDiff} symmetries. 
    
    In doing so, tree-level amplitudes are discussed in \cref{sec:trees} (finding complete agreement with GR), followed by a brief discussion on the one-loop formulation of Unimodular Gravity. 
    
    The main result, first found in \cite{Alvarez2016-2} and explicitly shown here, is that the Cosmological Constant is {\em stable under radiative corrections}. 
    
    Some important topics that have not found a place in this work are the construction of a hamiltonian for Unimodular Gravity, which can be used, for example, to study the Noether charges of the theory, an explicit discussion of the conformal factor in the path integral and the York-Gibbons-Hawking boundary correction.  
  \section{Acknowledgements.}
  We would like to thank our former unimodular collaborators; Jesús Anero, Diego Blas, Jaume Garriga, Sergio González-Martín, Mario Herrero-Valea, Juan José Lopez-Villarejo, Carmelo Pérez Martín, Raquel Santos-García, Enric Verdaguer, and Roberto Vidal.
  We also thank Ilya L. Saphiro for the careful revision of the manuscript and insightful comments.
  Authors acknowledge partial financial support by the Spanish MINECO through the Centro de Excelencia Severo Ochoa Program under Grant CEX2020-001007-S funded by MCIN/AEI/10.13039/501100011 033.
 Both authors acknowledge the European Union's Horizon 2020 research and innovation program under the Marie Sklodowska-Curie grant agreement No 860881-HIDDeN and also by Grant PID2019-108892RB-I00 funded by MCIN/AEI/ 10.13039/501100011033 and by "ERDF A way of making Europe."
 This project has received funds/support from the European Union's Horizon Europe program under Marie Sklodowska-Curie Actions-Staff Exchanges (SE) grant agreement No 101086085-ASYMMETRY.  

\end{document}